# Measuring and improving community resilience: a Fuzzy Logic approach

Melissa De Iuliis[a], Omar Kammouh[b], Gian Paolo Cimellaro[c]

[a] PhD. Candidate, Dept. of Structural, Geotechnical and Building Engineering, Politecnico di Torino, Corso Duca degli Abruzzi, 24, Torino, Italy, Email: melissa.deiuliis@polito.it

[b] PostDoc, Dept. Materials, Mechanism, Management and Design, Delft University of Technology, Stevinweg 1, 2628 CD Delft, Netherlands, Email: o.kammouh@tudelft.nl

[c] Associate Professor, Dept. of Structural, Geotechnical and Building Engineering, Politecnico di Torino, Corso Duca degli Abruzzi, 24, Torino, Italy (Corresponding author), Email: gianpaolo.cimellaro@polito.it

**Abstract**

Due to the increasing frequency of natural and man-made disasters worldwide, the scientific community has paid considerable attention to the concept of resilience engineering in recent years. Authorities and decision-makers, on the other hand, have been focusing their efforts to develop strategies that can help increase community resilience to different types of extreme events. Since it is often impossible to prevent every risk, the focus is on adapting and managing risks in ways that minimize impacts to communities (e.g., humans and other systems). Several resilience strategies have been proposed in the literature to reduce disaster risk and improve community resilience. Generally, resilience assessment is challenging due to uncertainty and unavailability of data necessary for the estimation process. This paper proposes a Fuzzy Logic method for quantifying community resilience. The methodology is based on the PEOPLES framework, an indicator-based hierarchical framework that defines all aspects of the community. A fuzzy-based approach is implemented to quantify the PEOPLES indicators using descriptive knowledge instead of hard data, accounting also for the uncertainties involved in the analysis. To demonstrate the applicability of the methodology, data regarding the functionality of the city San Francisco before and after the Loma Prieta earthquake are used to obtain a resilience index of the Physical Infrastructure dimension of the PEOPLES framework. The results show that the methodology can provide good estimates of community resilience despite the uncertainty of the indicators. Hence, it serves as a decision-support tool to help decision-makers and stakeholders assess and improve the resilience of their communities.

# 1 Introduction

Past global disaster events have shown an upward trend over the years, suggesting that modern communities are often not resilient enough to natural and man-made disasters. Therefore, research on disaster resilience has gained increased attention. Since resilience is a multidisciplinary concept and encompasses different research areas, several definitions of resilience can be found in the literature. The term resilience was introduced by Allenby and Fink [1] as "the ability of a system to remain in a practical state and to degrade gracefully in the face of internal and external changes". Bruneau, Chang [2], and later Cimellaro, Reinhorn [3], defined resilience as "the ability of social units to mitigate hazards, contain the effects of disasters when they occur, and carry out recovery activities to minimize social disruption and mitigate the effects of future earthquakes".

Disaster resilience is often classified into *technological units* and *social systems* [4]. The literature offers state-of-the-art approaches to quantify community resilience [5-10], most of which are indicator-based approaches. Among the available indicator-based resilience frameworks is the Hyogo Framework for Action (HFA) [11] [12], which is an internationally agreed top-down framework to increase the resilience of nations and communities through the implementation of detailed measures at the government and policy levels. Based on the Hyogo Framework, Kammouh, Dervishaj [13] have introduced a quantitative method to quantify resilience at the country level. Another top-down resilience framework is the Baseline Resilience Indicator for Communities (BRIC) [14], which is a quantitative framework that focuses on the inherent resilience of communities. A qualitative framework that measures resilience along with the ability to recover from seismic events is the San Francisco Planning and Urban Research Association framework (SPUR) [15]. It considers the recovery of buildings, infrastructures, and services to determine the resilience of physical infrastructure. Another hierarchical framework for evaluating community-level resilience was proposed by [16]. The model consists of community dimensions and their relationships with community services, systems, and resources. However, natural resources, which are an important indicator in the resilience planning process, were not considered in the proposed framework. Cimellaro, Renschler [17] presented the PEOPLES framework, which is a theoretical top-down framework that addresses all aspects of a community. These aspects are classified under seven community dimensions: Population; Environment; Organized government services; Physical infrastructure; Lifestyle; Economic; and Social capital. Later, the PEOPLES framework was upgraded into a quantitative framework for measuring community resilience [18, 19]. Nevertheless, the proposed methodology by the authors can only be implemented if the indicators and the relationships between them can be easily quantified, which may not be the case in certain scenarios where data collection comprises uncertainties and a lack of knowledge. Examples of bottom-up approaches to quantify community resilience to natural hazards include the Conjoint Community Resilience Assessment Measure (CCRAM) [20], the Communities Advancing Resilience Toolkit (CART) [21], and the Community Resilient System [22].

Despite this robust literature, there is still considerable disagreement about the indicators that define resilience and the frameworks that are most useful for measuring it. The scientific community is aware that the availability of data to be implemented in community resilience frameworks is one of the main issues. That is, the modeling approaches presented require accurate data to be incorporated into the models to be functional. Nevertheless, access to this data is limited and often the accuracy is insufficient. Furthermore, when unexpected events occur and there is not enough information or the previously established plan is not adequate, decisions are subjective and based on experience. The difficulty in the data and indicators acquisition process, as well as in defining the interaction between them, makes resilience assessment so complex that it cannot be used by decision-makers and industry. To respond to this challenge, many studies have focused on developing methods for quantifying community resilience and assessing the impacts of recovery strategies through probabilistic approaches, such as Bayesian Networks, to deal with stochastic uncertainties that can affect resilience assessment. For example, Schultz and Smith [23] developed a Bayesian network-based approach to evaluate the resilience of infrastructure networks and buildings in Jamaica Bay, New York. Kammouh, Gardoni [24] introduced a novel approach to assess the time-dependent resilience of engineering systems using resilience indicators through the Dynamic Bayesian Network (DBN). Another Bayesian network-based approach for seismic resilience quantification was proposed by [25]. Cai, Lam [26] employed a Bayesian network to investigate interdependencies of resilience components and improve disaster resilience. Furthermore, Kameshwar, Cox [27] developed a probabilistic decision support framework for community resilience planning under multiple hazards using Bayesian Network. Despite the advantages of the Bayesian network, such as updating the system to which it is applied when new data and information become available, the main concern is its application in case of epistemic uncertainties and the computational effort in determining conditional probabilities [28].

The primary goal of this paper is to cover the previously mentioned shortcomings of existing scientific literature by introducing a Fuzzy Logic-based method within the context of the PEOPLES framework. The proposed method utilizes the resilience indicators presented in [19] to come up with an extensive resilience model that accounts for all aspects of a community. The fuzzy logic technique is used for inference to account for the data-related uncertainties. The methodology here derived does not require precise and deterministic data for its implementation to assess the resilience of urban communities.

The contributions of this work are summarized as follows:

1. Developing an extensive and comprehensive indicator-based model that captures all fundamental aspects of a community.
2. Introducing a weighting technique that ranks the indicators according to their importance.
3. Employing the fuzzy logic inference technique to account for data uncertainties of the analyzed indicators.
4. Presenting a case study to demonstrate the applicability of the introduced resilience estimation methodology.
5. Verifying the methodology by comparing the model output with the output obtained from [19].

The resilience quantification methodology presented in this paper can be used as a decision-support tool by decision-makers to (i) learn about the state of their communities in the face of a particular event and (ii) prioritize planning and management strategies to improve the resilience of their communities to future hazardous events. The remainder of the paper is organized as follows. Section 2 reviews the PEOPLES framework along with its seven dimensions. Section 3 is dedicated to reviewing the basic knowledge of the Fuzzy Logic and its implementation within the PEOPLES framework. Section 4 describes the proposed methodology for estimating community resilience. Section 5 presents a demonstrative example to demonstrate and verify the proposed resilience estimation model. Finally, conclusions are drawn in Section 6 together with the proposed future work.

# 2  PEOPLES Framework

PEOPLES is a multi-layered framework developed at the Multidisciplinary Center of Earthquake Engineering Research (MCEER) that aims to identify different resilience characteristics of a community at different scales (spatial and temporal) and assess possible responses of a community by taking into account the interdependence between community levels [5]. The PEOPLES framework consists of seven dimensions of a community divided into a set of components, each of which is in turn subdivided into several indicators. The seven dimensions are summarized by the acronyms PEOPLES as follows [4, 29] (see Figure 1):

1. D1-Population and Demographics: considers the socioeconomic composition of the community and measures social vulnerabilities that could affect the emergency response and recovery systems (e.g., minority and socioeconomic status, age distribution, population density).

2. D2-Environment and Ecosystem: estimates the ability of the ecosystem and environment to return to pre-hazard conditions.

3. D3-Organized Government Services: concerns the community services that the government guarantees before and after a hazard event. This includes preparedness and mitigation strategies (before the event) and response and recovery plans (after the event).

4. D4-Physical Infrastructure: takes into account the facilities and lifelines that must be restored to their initial condition after a hazard event.

5. D5-Lifestyle and Community Competence: defines a community's capabilities (e.g., the ability to face complex problems and find appropriate solutions employing political networks) and perceptions (e.g., the judgments and feelings a community has about itself toward a positive change).

6. D6-Economic development consists of the current economic state (static state) of the community and its development and future growth (dynamic state).

7. D7-Social-cultural capital evaluates the community's attitude to bounce back to the pre-event conditions.

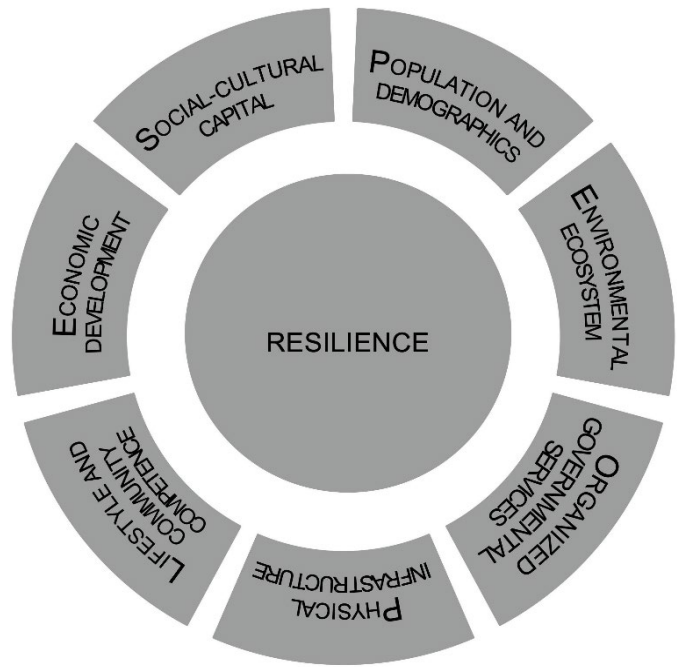

Figure 1. PEOPLES Resilience Framework - Dimensions

Every dimension of the PEOPLES framework is divided into a set of components and every component is further broken down into a set of indicators. This expansion of the dimensions is performed to capture all aspects of a community. Regarding the indicators, a list of 115 resilience indicators found in the literature was collected and allocated to the proper components of PEOPLES [19]. Each indicator has been made computable by assigning it a measure that allows the analytical evaluation of the indicator's performance. Each measure is normalized between 0 and 1 to a fixed quantity known as the target value (*TV*). The target value provides the baseline for measuring the resilience of a system [30]. Furthermore, the measures are classified into "static measures (S)" and "dynamic measures (D)". A static measure is a measure that is not impacted by a hazardous event, while a dynamic measure is a measure that is event-sensitive (i.e., the value of the measure changes following a hazard event). The variables (i.e., dimensions, components, and indicators) included in the PEOPLES framework do not contribute equally to the resilience output. Therefore, they are classified according to their importance. Each variable in the same group is assigned an importance factor (I) which is normalized using a min-max rescaling technique. The min-max rescaling technique is used to scale each variable between 0 and 1, where 0 corresponds to the worst rank and 1 represents the best rank. To represent the functionality of each variable within the PEOPLES framework, a set of parameters obtained from past events or by performing hazard analysis is used: un-normalized initial functionality $q_{0u}$, normalized initial functionality before the event $q_0$, post-disaster functionality $q_1$, the functionality after recovery $q_r$, and the restoration time $T_r$ required to complete the recovery process [19].

The PEOPLES framework has also been implemented as an online tool (http://www.resiltronics.org/PEOPLES/logic.php). After providing all necessary data, the tool evaluates the Loss of Resilience (*LOR*) of the analyzed community. The full description of the tool can be found in [31].

## 3  Background of Fuzzy Logic

Zadeh [32] introduced the concept of fuzzy set theory and fuzzy logic to address the subjectivity of human judgment in the use of linguistic terms in the decision-making process [33-35]. The purpose of fuzzy logic is to solve high degree uncertainty problems and to represent vague, ambiguous, and chaotic information [36, 37]. Over the years, Fuzzy Logic has become a key factor in many fields due to its effectiveness and reliability.

In the existing literature, fuzzy set theory and fuzzy logic have been applied in Machine Intelligence Quotient (MIQ) to simulate human ability, in earthquake engineering for seismic damage evaluation [35, 38], in fragility curve analysis [39], and natural disaster risk management [40].

Fuzzy logic assigns different membership grades ($\mu$) ranging between 0 and 1 to a variable *x* to indicate the membership of the variable to several classes (fuzzy sets). The strength of the fuzzy logic inference system relies on the following two main features: (i) fuzzy inference system can handle both descriptive (linguistic) knowledge and numerical data; (ii) fuzzy inference system uses approximate reasoning algorithm to determine relationships between inputs by which uncertainties can be propagated throughout the process [41]. Implementing fuzzy logic as an inference system to quantify the resilience, requires three main steps: 1) fuzzification and membership functions; 2) Fuzzy Inference System (FIS) to aggregate the indicators, and 3) defuzzification (Figure 3).

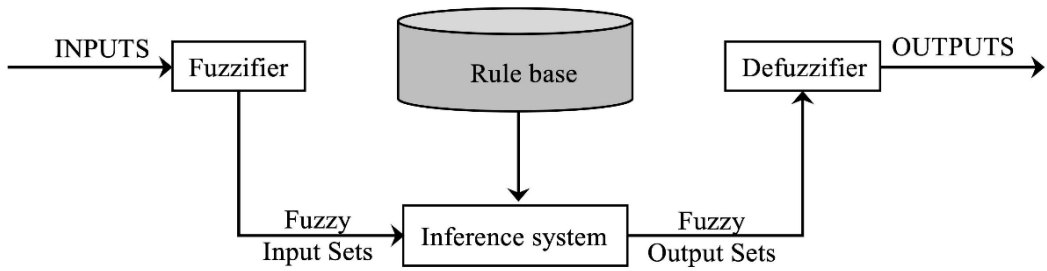

Figure 2. Fuzzy Inference System (FIS)

# 4 Methodology to estimate community resilience

The methodology proposed in this work enhances the previous work introduced by [19] by incorporating fuzzy logic in the computation process. The methodology can be divided into the following (see Figure 2):

- Resilience modeling and indicator grouping: a hierarchical rule base model is built based on the structure of the PEOPLES framework. Indicators belonging to a specific component are further divided into subgroups according to predefined criteria.

- Interdependency analysis and importance factor: weighting factors and importance factors are allocated to each PEOPLES variable (i.e., indicators, components, and dimensions) as they do not contribute equally to the overall resilience output.

- Inference: the last step of the methodology is to combine the indicators through an FL inference system and obtain the final output of the framework.

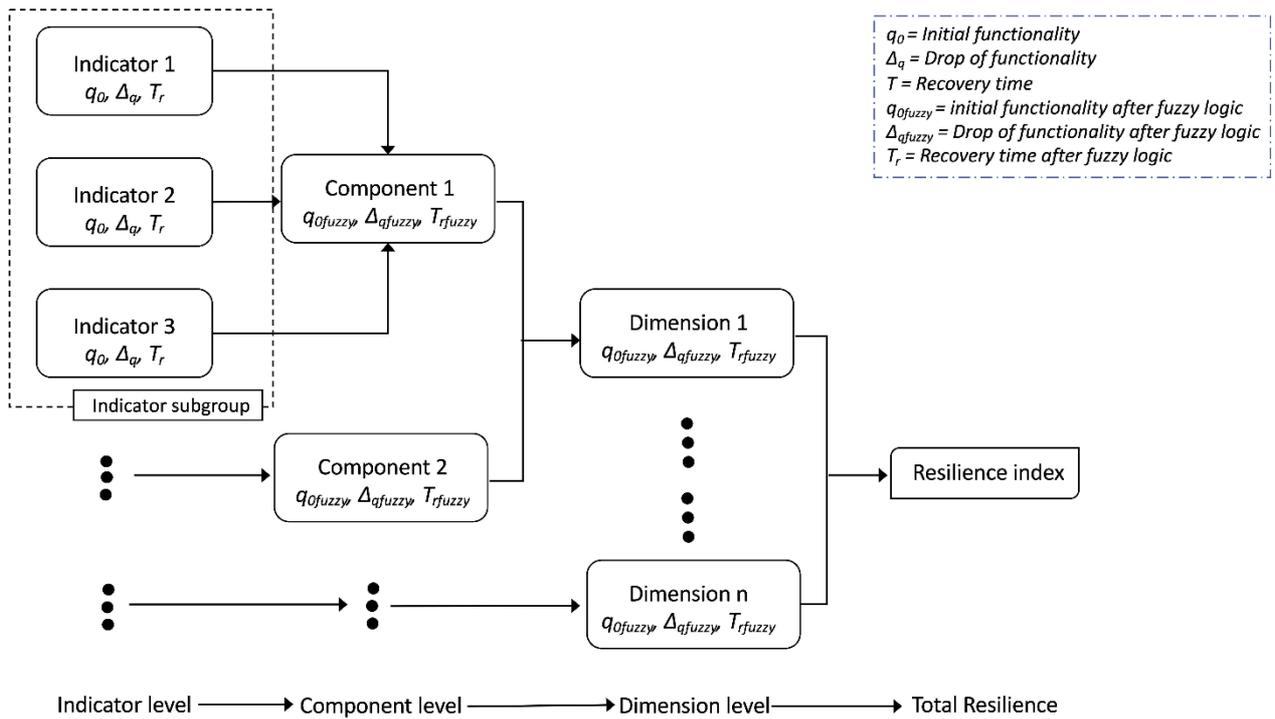

Figure 3. Hierarchical rule base model applied to PEOPLES framework

## 4.1 Step 1: Resilience modeling and indicators grouping

The first step of the methodology is the definition of a hierarchical scheme for the seven dimensions of the PEOPLES framework. A total of eight flowcharts are presented (figures Figure 4-Figure 11); i.e., seven flowcharts for the seven dimensions and an additional flowchart for the final resilience output. Indicators within each component are further clustered into subgroups with no more than 3 indicators each. Indicators that belong to the same subject or field are clustered in one group. This additional grouping of indicators allows having simple hierarchical relationships between the indicators and the components for the subsequent implementation of the fuzzy logic.

*Population and demographics*

This dimension contains indicators that describe the population and demographics in a given community. For instance, *median income* and *age distribution* information is essential to measure the economic health of the community. The *Population and demographics* dimension comprises 3 components, *Distribution/density, Composition*, and *Socioeconomic status*, having a total of 9 indicators. Indicators within this dimension are clustered into 6 subgroups, namely, Percentage of population, Family asset, Economic diversity, Aggregation,

Population equity, Disparities, and Demography. The hierarchical scheme designed for the *Population and demographics* dimension is shown in Figure 4.

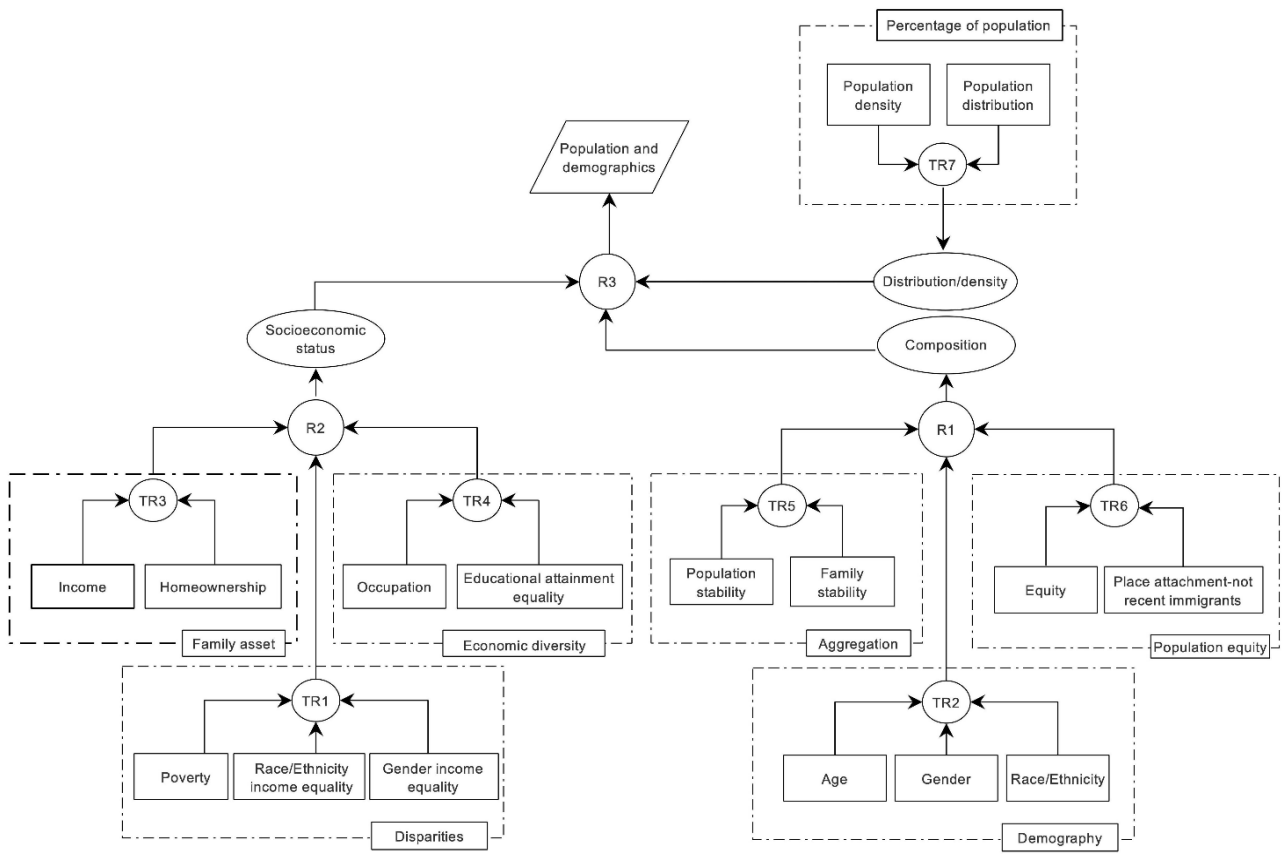

Figure 4. *Population and demographics* dimension hierarchical scheme

*Environment and ecosystem*

The *Environment and ecosystem* dimension measures the ability of the ecological system to restore its pre-event state. The *Environment and ecosystem* dimension contains 6 components: *Water, Air, Soil, Biodiversity, Biomass (Vegetation)*, and *Sustainability*, with a total of 13 indicators. Indicators within this dimension are classified into 5 subgroups: Environment quality, Percentage of land, Land type, Land use, and Vegetation index (see Figure 5). Note that components having a single indicator are processed as if they are a single component. In other words, the indicators of those components are clustered within the same subgroup.

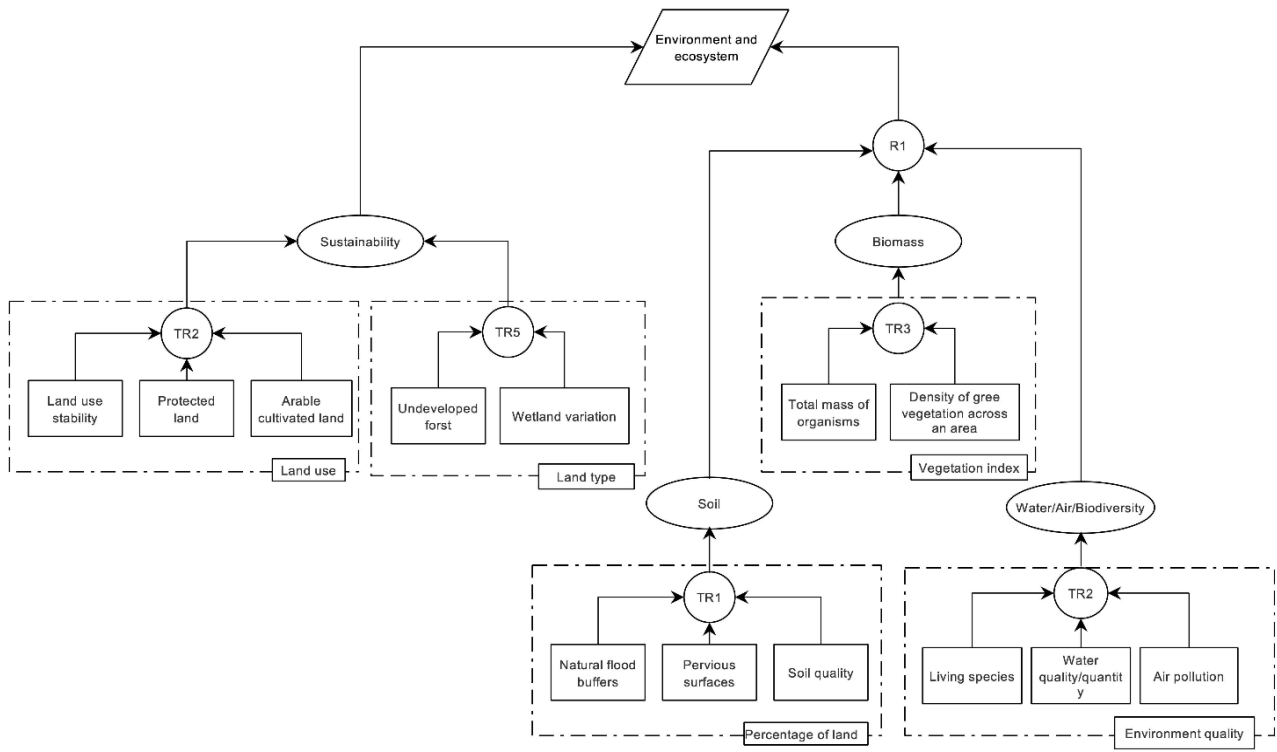

Figure 5. *Environment and ecosystem* dimension hierarchical scheme

*Organized governmental services*

The *Organized governmental services* dimension includes information about traditional legal and security services such as police, emergency, and fire departments as well as services provided by public health, hygiene departments, and cultural heritage departments. The indicators within this dimension are also related to disaster emergency plans, training, and other operations that might help ensure proper disciplined responses. The *Organized governmental services* dimension comprises 5 components: *Executive/administrative*, *Judicial*, *Legal/security*, *Mitigation/preparedness*, and *Recovery/response*, with a total of 26 indicators. As the *Judicial* component presents one indicator, it has been linked to indicators of the *Executive/administrative* component. As shown in Figure 6, the hierarchical scheme consists of 10 subgroups where indicators are aggregated to get the final result of the *Organized governmental services* dimension.

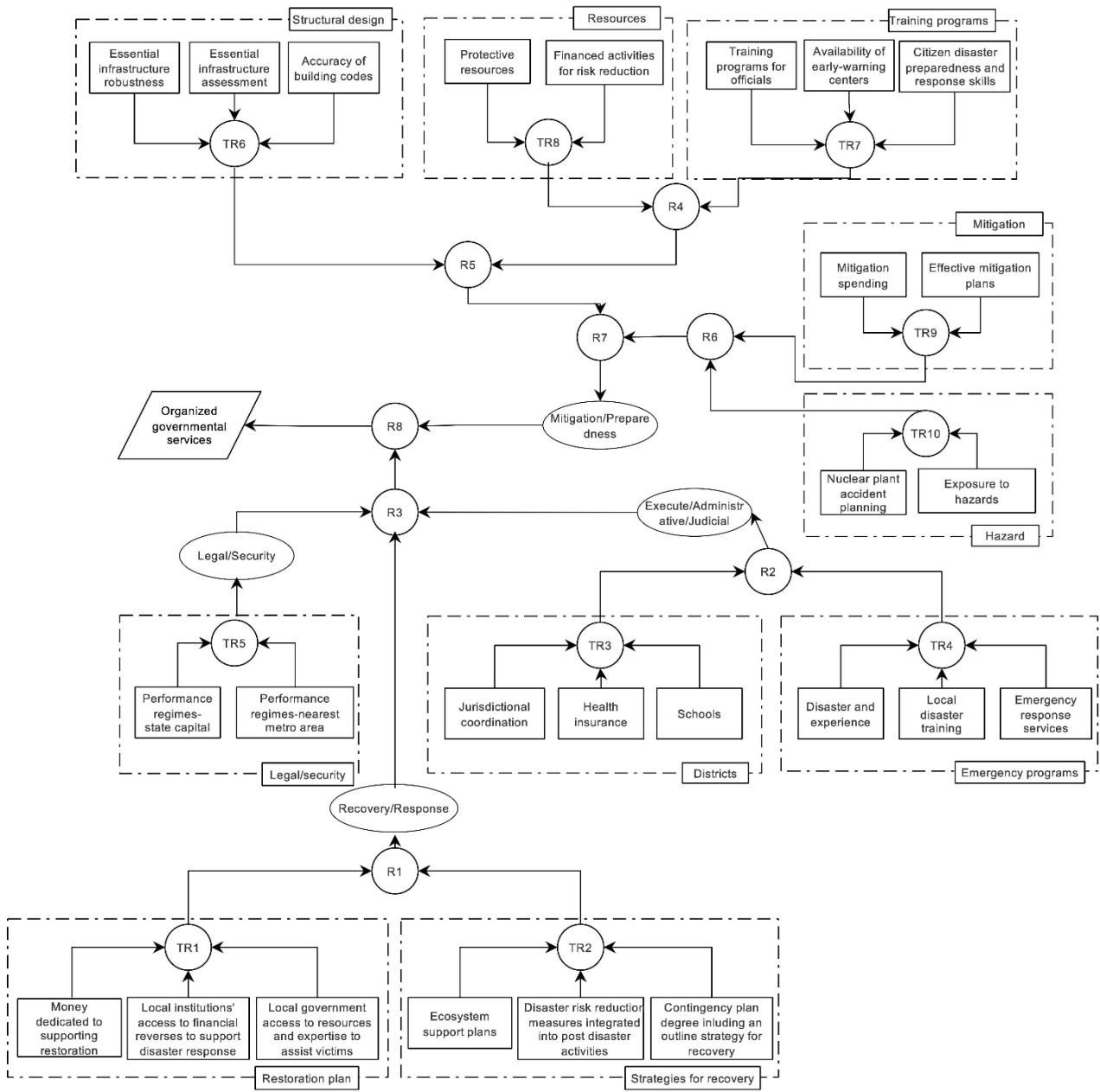

Figure 6. *Organized governmental service* dimension hierarchical scheme

*Physical infrastructure*

The *Physical infrastructure* dimension emphasizes the built environment of a community. It incorporates both *Facilities* and *Lifelines* components with a total of 21 indicators, as illustrated in Figure 7. Indicators included within the *Facilities* component refer to housing, commercial and cultural facilities. Indicators under the *Lifelines* component consider food supply, health care, utilities, transportation, and communication networks. The hierarchical scheme is structured in 7 subgroups: Communication, Evacuation, Healthcare, Services, Commercial activities, Housing, and Supply.

Among PEOPLES dimensions, *Physical infrastructure* is often the dimension that needs urgent attention in the aftermath of a hazardous event. That is, authorities and government services work to restore the utilities' functionality to allow critical facilities to perform their functions.

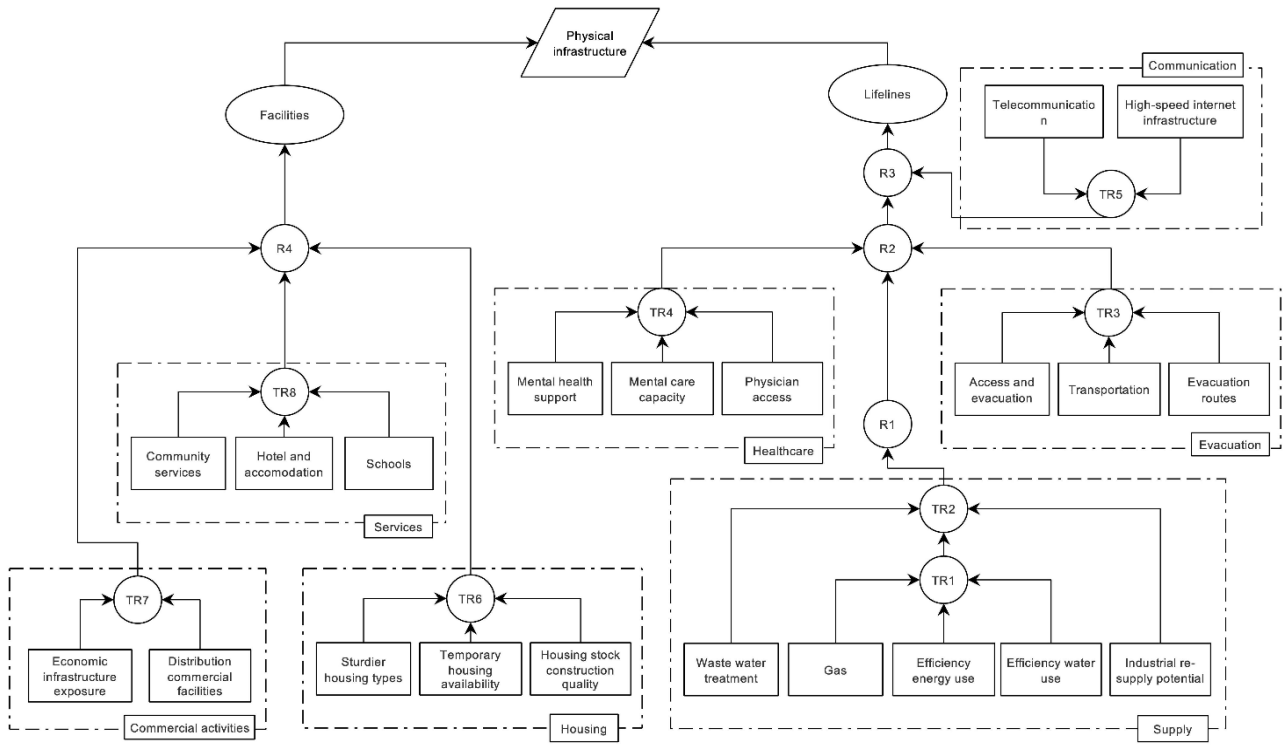

Figure 7. *Physical Infrastructure* dimension hierarchical scheme

*Lifestyle and community competence*

This dimension deals with the ability of a community to develop solutions to complex problems including warning plans, procedures, and disaster training programs. Also, the participation of community members in the activities required to sustain the community is a key indicator to measure community resilience. It is believed that communities that collectively believe that they can face complex problems are more likely resilient against environmental and governmental obstacles.

*Lifestyle and community competence* is divided into 2 components: *Quality of life* and *Collective actions and efficacy*, with a total of 7 indicators. The hierarchical framework of the *Lifestyle and community competence* dimension is organized into 3 subgroups: Ability, Security, and Neighborhood (see Figure 8).

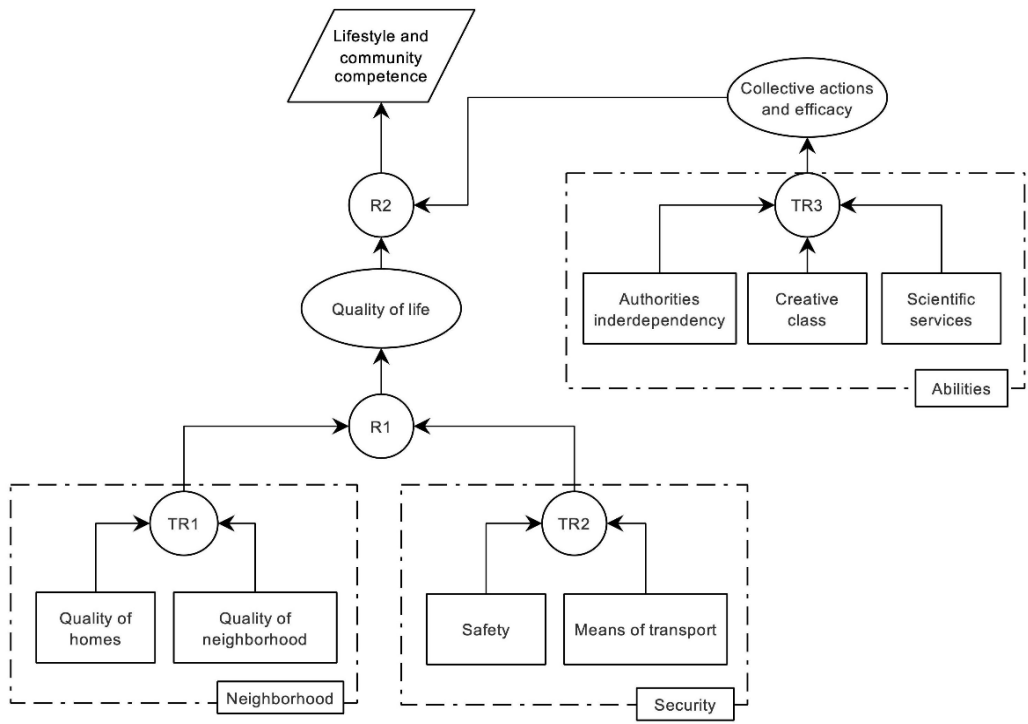

Figure 8. *Lifestyle and community competence* dimension hierarchical scheme

*Economic development*

The *Economic development* dimension includes both the static evaluation of a community's current economy (economic activity) and the dynamic evaluation of a community's economic growth (economic development). The former takes into account the provision of labor for the production of economic goods and services, the latter measures a community's productive capacities in terms of technologies, technical cultures, and the capacities and skills of those engaged in production. Key indicators of the *Economic development* dimension focus also on life expectancy and poverty level. Thus, it is evident that this dimension is closely connected with the *Population and Demographics* dimension. Other key indicators cover the availability of evacuation plans and drills for structures, adequacy plans for damaged buildings, and commercial reconstruction following a disaster. The *Economic development* dimension consists of 3 components: *Financial services*, *Industry production*, and *Industry employment services*, with a total of 16 indicators. 7 subgroups are used to grouped indicators in the hierarchical framework, as illustrated in Figure 9.

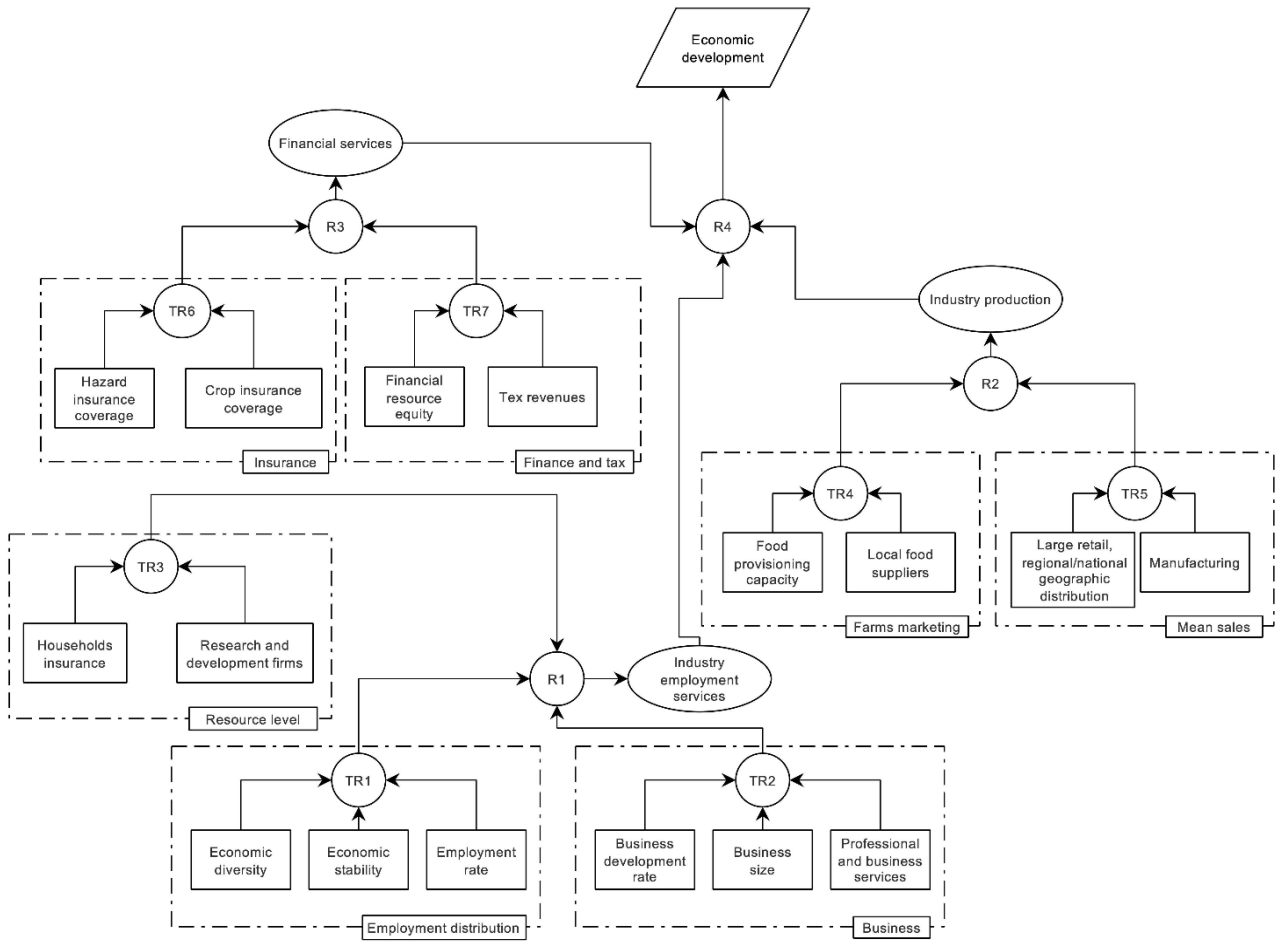

Figure 9. Economic development dimension hierarchical scheme

*Social-cultural capital*

The last dimension of the PEOPLES framework, namely *Social-cultural capital,* incorporates education, social, and cultural services, child and elderly services, and community participation in formal organizations such as religious congregations, schools, and resident associations. The *Social and cultural capital* dimension is measured through the acquisition of surveys concerning the number of members of civil and community organizations, and the quality of life. Furthermore, key indicators include the existence of adequacy plans and management plans following a hazard.

The *Social-cultural capital* dimension is splitting into 7 components: *Community participation*, *Nonprofit organization*, *Place attachment*, *Child and elderly activities*, *Commercial centers*, *Cultural and heritage services*, and *Education services*, with a total of 17 indicators (see Figure 10). Indicators within this dimension are classified into 6 subgroups: Social organizations, Participation classes, Participation, Social and civic programs, Risk reduction programs, and Education.

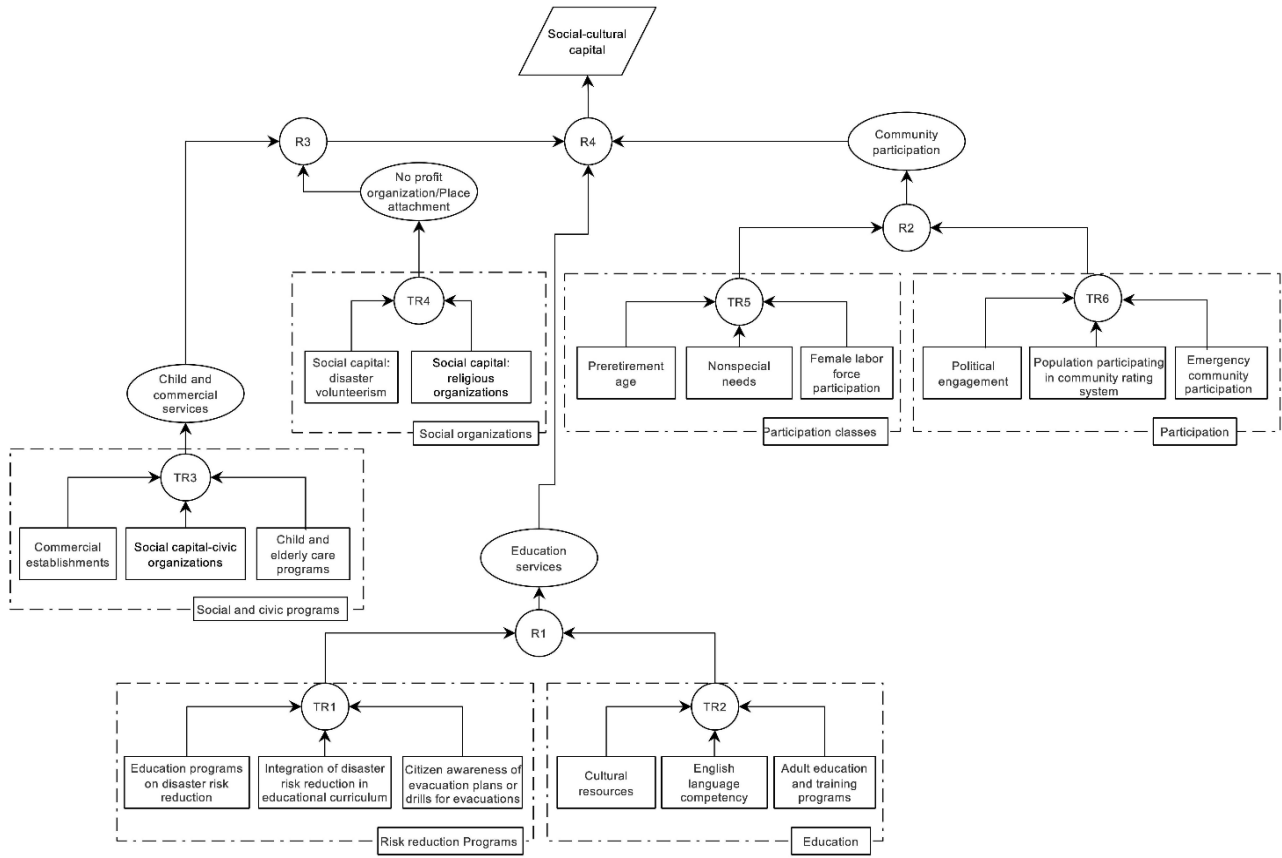

Figure 10. *Social-cultural capital* dimension hierarchical scheme

*The final Resilience output*

The final hierarchical scheme that combines the seven dimensions is shown in Figure 11. The *Population and demographics* dimension is closely related to the *Economic development* dimension as the latter involves the life expectancy and poverty rates of the population, and therefore they are grouped. The *Physical infrastructure* dimension is largely related to the *Organized governmental services* dimension, which considers the infrastructure robustness and assessment, and the availability of resources for recovery programs. The *Environment and ecosystem* dimension depends on the *Organized governmental services* dimension, which verifies the availability of local government plans to support the restoration, protection, and sustainable management of ecosystem services [42]. Finally, the *Social-cultural capital* dimension is considered a prerequisite to *Lifestyle and community competence* as the *Social-cultural capital* dimension incorporates different services that a community has provided for itself [43].

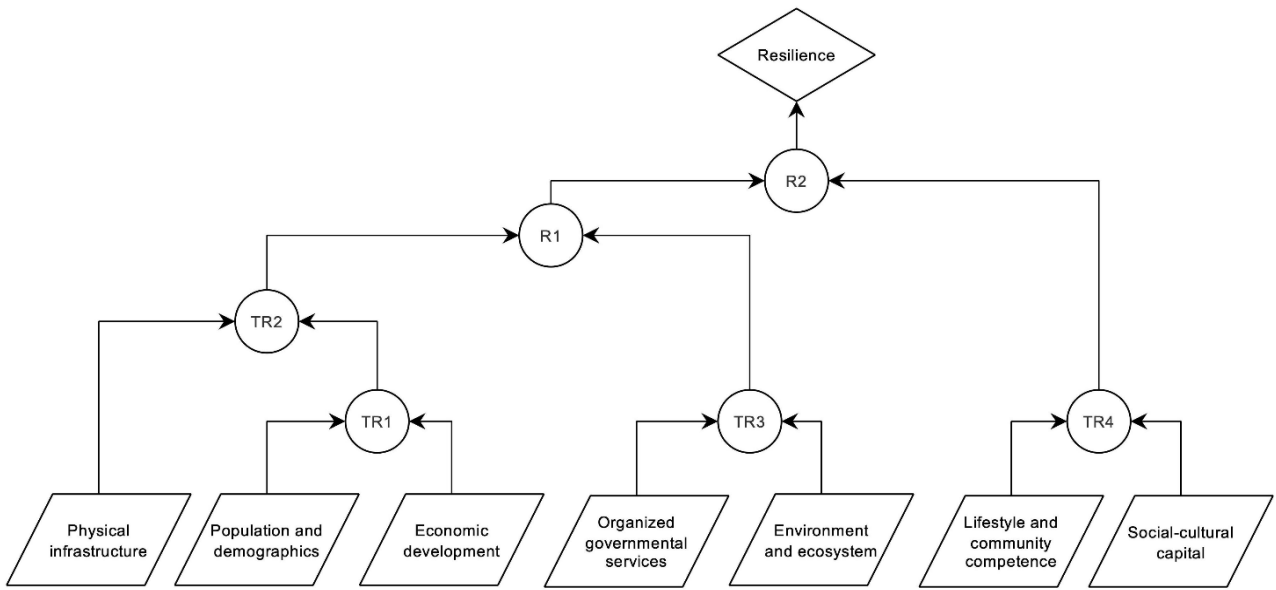

Figure 11. Resilience hierarchical scheme

## 4.2 Step 2: Interdependency analysis and importance factor

PEOPLES indicators do not contribute equally to the overall resilience outcome; hence, the interdependencies among them can affect the final result. To include interdependencies in this work, weighting factors are assigned to each variable through an interdependency analysis. In the analysis, variables of the PEOPLES framework are classified into three main groups:

1. Indicators within a component are considered as a group (29 groups in total);
2. Components that fall within a dimension are takes as a group (7 groups in total);
3. The seven dimensions of PEOPLES make up a group (1 group).

The interdependency analysis assumes that the importance of a variable is related to the number of other variables in the same group that depends on it. Variables in the same groups are given importance factors using the [$n \times n$] adjacency matrix in Figure 12, where $n$ is the number of variables in the analyzed group. Each cell ($a_{ij}$) in the matrix represents the degree of dependence between two variables and can take the values 0 or 1. A value of 0 indicates that the variable in the row does not depend on the variable in the column, while 1 indicates that the variable in the row depends entirely on the variable in the column. The importance factor is carried out by summing up the values in each column of the matrix.

An interdependency matrix is built for each group of variables. That is, a single interdependency matrix is constructed for the seven dimensions of PEOPLES, for each group of components under the dimensions, and finally for every group of indicators under the components. This results in 37 matrices to perform the interdependency analysis for the different variables of the PEOPLES framework.

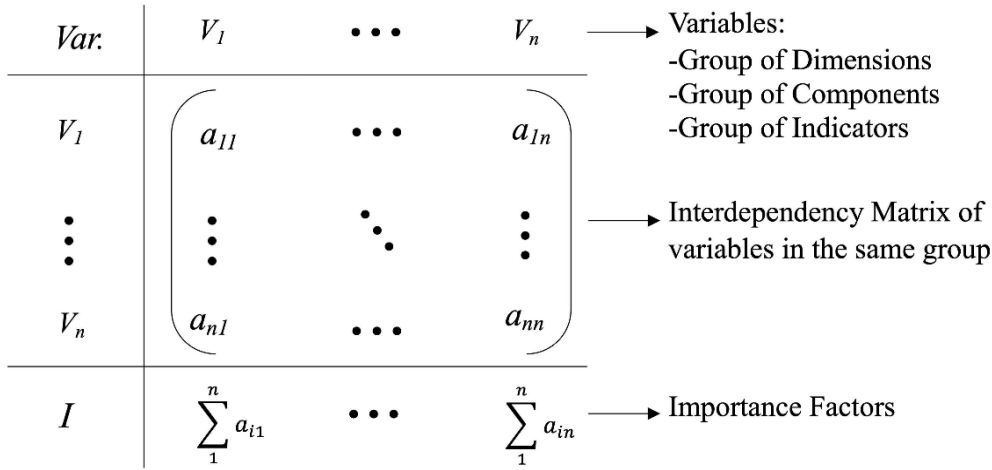

Figure 12. Interdependency matrix between variables in a same group

The level of interdependency between two variables can be identified using descriptive knowledge in the form of a walk-down survey filled by a team of experts. For instance, "low" and "high" dependence between two variables can be translated into 0 and 1, respectively. Finally, a weighting factor for each variable ($w_i$) is obtained by dividing the importance factor by the maximum importance factor, as indicated in Equation 2:

$$w_i = \frac{\sum_{j=1}^{n} a_{ji}}{\sum_{i=1}^{n}\sum_{j=1}^{n} a_{ji}} \tag{1}$$

## 4.3 Step 3: Inference system

The last step of the methodology is the implementation of the fuzzy logic inference system for combining the PEOPLES variables to get the resilience index. As described in Section 3.1, applying fuzzy logic as an inference system to estimate community resilience requires three major steps: (i) fuzzification and definition of membership functions, (ii) determining fuzzy rule tables to aggregate PEOPLES variables, and (iii) defuzzification. In the following, the fuzzy logic process is described.

### 4.3.1 Fuzzification process – Membership functions

The fuzzification process is helpful when it comes to uncertainties in the estimation of system inputs as it can cope with both numerical and descriptive variables. In the case of numerical input, the fuzzification process is straightforward and depends on the shape of the membership functions. That is, membership functions can have different forms, such as triangular, trapezoidal, and Gaussian shapes. Descriptive inputs, instead, can be directly converted in fuzzy terms by assigning different membership degrees to the different granularities. For example, within the PEOPLES framework, if the average number of internet connections, television, radio, and telephone per household in a community is classified as "poor", the indicator *Telecommunication* can be converted in fuzzy terms by assigning the membership degrees to the granularities [L, M, and H] as follows: [$\mu_L$, $\mu_M$, $\mu_H$] = [0.9, 0.1, 0]. The chosen membership degrees show a low level of functionality for the *Telecommunication* indicator.

### 4.3.2 Fuzzy Inference System (FIS) – fuzzy rules

A Fuzzy Inference System (FIS) aims at mapping the input information into an output space exploiting the fuzzy sets previously defined. The relationships between inputs and outputs are defined through the *fuzzy rule base* (FRB) that comes from the heuristic knowledge of experts or historical data. In this work, the Mamdani fuzzy inference system, which consists of if-then statements (rules) that link the input (antecedent) to a consequent (output), is adopted. In a complex system with a large number of input indicators, the number of rules must cover all the possible combinations of them.

### 4.3.3 Defuzzification process – crisp number

The defuzzification step is carried out to obtain a crisp number from the output fuzzy set resulting from the inference process. The defuzzification is performed according to the membership function of the output variable and represents the inverse of the fuzzification process. Several defuzzification techniques have been developed, such as the center of gravity, center of area, and mean of maximum method [44].

# 5 Demonstrative example and verification

To demonstrate the applicability of the proposed fuzzy logic-based methodology, the resilience of the city of San Francisco is evaluated. The 1989 Loma Prieta earthquake, which was characterized by a moment magnitude of 6.9 $M_w$, is considered the hazard event. The introduced methodology has been used to carry out the case study, focusing only on the *Physical Infrastructure* dimension. Therefore, in this example, the resilience outcome is obtained by taking into account only the contribution of the *Physical Infrastructure* dimension. It includes facilities (e.g., housing, commercial and industrial facilities, and cultural facilities) and lifelines (food supply, utilities, transportation, communication networks) within a built environment [45]. Table 1 shows the list of indicators and components along with the corresponding functionality and repair time parameters that are required to quantify the resilience. For this study, open database sources were investigated to determine the parameters of the San Francisco community [46]. In Table 1, $q_{0u}$ is the un-normalized initial functionality that must be normalized to be combined with the other parameters. The normalization of the initial functionality $q_0$ is done by dividing the un-normalized functionality $q_{0u}$ over the Target Value *TV* described before. Furthermore, the recovery time parameter $T_r$ is normalized based on a 3-year time span, which is normally the time reference for civil applications.

Table 1. Functionality parameters of indicators within Physical Infrastructure dimension of the San Francisco city. The reported values are retrieved from [19]

| Dimension/component/ subgroups/indicators | Measure | Nat. | $q_{0u}$ | TV | $q_0$ | $q_1$ | $\Delta q$ | $T_r$ |
|---|---|---|---|---|---|---|---|---|
| **4 - *Physical infrastructure*** | | | | | | | | |
| **4-1 - *Facilities*** | | | | | | | | |
| **Housing** | | | | | | | | |
| 4-1-1 - Sturdier housing types | % housing units not manufactured homes | D | 1 | 1 | 1 | 0,599 | 0,401 | 120 |
| 4-1-2 - Temporary housing availability | % vacant units that are for rent | D | 2,68 | 5 | 0,536 | 0,05 | 0,486 | 620 |
| 4-1-3 - Housing stock construction quality | 100-% housing units built prior to 1970 | D | 0,241 | 1 | 0,241 | 0,145 | 0,096 | 700 |
| **Commercial Activities** | | | | | | | | |
| 4-1-4 - Economic infrastructure exposure | % commercial establishments outside of high hazard zones ÷ total commercial establishment | S | 0,85 | 1 | 0,850 | 0,850 | 0 | - |
| 4-1-5 - Distribution commercial facilities | % commercial infrastructure area per area ÷ TV | D | 0,3 | 0,15 | 0,867 | 0,520 | 0,347 | 160 |
| **Services** | | | | | | | | |
| 4-1-6 - Community services | % Area of community services (recreational facilities, parks, historic sites, libraries, museums) total area ÷ TV | D | 0,16 | 0,2 | 0,800 | 0,480 | 0,320 | 430 |
| 4-1-7 - Hotels and accommodations | Number of hotels per total area ÷ TV | D | 102 | 128 | 0,797 | 0,478 | 0,319 | 130 |
| 4-1-8 - Schools | Schools' area (primary and secondary education) per population ÷ TV | D | 134 | 140 | 0,957 | 0,574 | 0,383 | 90 |

4-2 - *Lifelines*

**Healthcare**

| | | | | | | | | |
|---|---|---|---|---|---|---|---|---|
| 4-2-1 - Mental health support | Number of beds per 100000 population ÷ TV | D | 69 | 75 | 0,920 | 0,644 | 0,276 | 35 |
| 4-2-2 - Medical care capacity | Number of available hospital beds per 100000 population ÷ TV | D | 544 | 600 | 0,907 | 0,635 | 0,272 | 35 |
| 4-2-3 - Physician access | Number of physicians per population ÷ TV | S | 2,5 | 3 | 0,833 | 0,833 | 0 | - |

**Evacuation**

| | | | | | | | | |
|---|---|---|---|---|---|---|---|---|
| 4-2-4 - Access and evacuation | Principal arterial miles per total area ÷ TV | D | 172138 | 2E+05 | 0,861 | 0,602 | 0,259 | 45 |
| 4-2-5 - Transportation | Number of rail miles per area ÷ TV | D | 5412 | 6000 | 0,902 | 0,631 | 0,271 | 72 |
| 4-2-6 - Evacuation routes | Major road agrees points per building ÷ TV | S | 0,67 | 1 | 0,670 | 0,670 | 0 | - |

**Supplies**

| | | | | | | | | |
|---|---|---|---|---|---|---|---|---|
| 4-2-7 - Efficient energy use | Ratio of Megawatt power production to demand | D | 1 | 1 | 1 | 0,240 | 0,760 | 25 |
| 4-2-8 - Efficient Water Use | Ratio of water available to water demand | D | 1 | 1 | 1 | 0,240 | 0,760 | 60 |
| 4-2-9 - Gas | Ratio of gas production to gas demand | D | 0,1 | 1 | 0,100 | 0,050 | 0,050 | 70 |
| 4-2-10 - Industrial re-supply potential | Rail miles per total area ÷ TV | D | 5412 | 6000 | 0,902 | 0,631 | 0,271 | 45 |
| 4-2-11 - Waste water treatment | Number of WWT units per population ÷ TV | D | 3 | 4 | 0,750 | 0,300 | 0,450 | 65 |

**Communication**

| | | | | | | | | |
|---|---|---|---|---|---|---|---|---|
| 4-2-12 - Telecommunication | Average number of internets, television, radio, telephone, and telecommunications broadcasters per household ÷ TV | D | 5 | 6 | 0,833 | 0,500 | 0,333 | 90 |
| 4-2-13 - High-speed internet infrastructure | % population with access to broadband internet service | D | 0,9 | 1 | 0,900 | 0,450 | 0,450 | 300 |

*Step 1: Resilience modeling and indicators grouping*

The first step of the proposed methodology is the definition of a hierarchical framework for the analyzed dimension. As illustrated in Figure 7, indicators together with the corresponding parameters belonging to *Facilities* and *Lifelines* components are divided into subgroups with no more than 3 indicators each. The Indicators are clustered in 7 subgroups following the PEOPLES structure: Housing, Commercial Activities, Services, Healthcare, Evacuation, Supplies, and Communication. In every subgroup, indicators (e.g., *telecommunication*, *high-speed internet infrastructure*, etc.) are combined through fuzzy rules to obtain *Facilities* and *Lifelines* components. Finally, the components in turn are combined to get the resilience output.

*Step 2: Interdependency analysis and importance factor*

Once the hierarchical framework for the studied dimension is built, the second step of the methodology starts. The weighting factors of the different variables under the *Physical infrastructure* dimension are determined using the interdependency matrix technique. As an example, the interdependency matrix of the indicators within the *Lifelines* component is shown in Table 2.

Table 2. Interdependency matrix between indicators under the Lifelines component

| | Internet infrastructure | Efficient energy use | Water use | Gas | Access and evacuation | Transport | Wastewater treatment |
|---|---|---|---|---|---|---|---|
| Internet infrastructure | — | 1 | 0 | 0 | 1 | 1 | 0 |
| Efficient energy use | 0 | — | 0 | 0 | 0 | 0 | 0 |
| Water use | 0 | 0 | — | 0 | 0 | 0 | 0 |
| Gas | 0 | 1 | 1 | — | 0 | 1 | 1 |
| Access and evacuation | 0 | 1 | 1 | 1 | — | 1 | 1 |
| Transport | 0 | 1 | 0 | 0 | 1 | — | 0 |
| Wastewater treatment | 1 | 1 | 1 | 1 | 1 | 1 | — |
| Sum | 2 | 11 | 7 | 5 | 8 | 9 | 6 |

| Indicator | Telecommunication | Mental health support | Physician access | Medical care capacity | Evacuation routes | Industrial resupply potential |
|---|---|---|---|---|---|---|
| Telecommunication | 1 | 0 | 0 | 0 | 0 | 0 |
| Mental health support | 0 | 1 | 0 | 1 | 0 | 0 |
| Physician access | 0 | 0 | 1 | 1 | 0 | 0 |
| Medical care capacity | 1 | 0 | 1 | 1 | 0 | 0 |
| Evacuation routes | 0 | 0 | 0 | 0 | 1 | 0 |
| Industrial resupply potential | 0 | 0 | 0 | 0 | 1 | 1 |
| Internet infrastructure | 1 | 0 | 0 | 0 | 0 | 1 |
| Efficient energy use | 0 | 0 | 0 | 0 | 0 | 0 |
| Water use | 1 | 0 | 0 | 0 | 0 | 0 |
| Gas | 1 | 0 | 0 | 0 | 0 | 0 |
| Access and evacuation | 1 | 0 | 0 | 0 | 1 | 0 |
| Transport | 1 | 0 | 0 | 0 | 1 | 0 |
| Wastewater treatment | 1 | 0 | 0 | 0 | 0 | 1 |
| Importance factor | 8 | 1 | 2 | 3 | 4 | 3 |

Once the importance factors have been extracted from the interdependency matrix, weighting factors for indicators and components under the Physical infrastructure dimension are obtained through Eq. (1). Table 3 lists the weighting factors of the different variables under the *Physical infrastructure* dimension.

Table 3. Weighting factors of variables within *Physical infrastructure* dimension for city of San Francisco

| Component/Indicator | w |
|---|---|
| 4.1 Facility | 0.5 |
| **Housing** | |
| 4.1.1 Sturdy housing types | 0.5 |
| 4.1.2 Temporary housing availability | 0.5 |
| 4.1.3 Housing stock construction quality | 0.75 |
| **Commercial activities** | |
| 4.1.4 Economic infrastructure exposure | 0.75 |
| 4.1.5 Distribution commercial facilities | 0.5 |
| **Services** | |
| 4.1.6 Community services | 1 |
| 4.1.7 Hotels and accommodations | 0.75 |
| 4.1.7 Schools | 0.5 |
| 4.2 Lifelines | 1 |
| **Healthcare** | |
| 4.2.1 Mental health support | 0.09 |
| 4.2.2 Medical care capacity | 0.27 |
| 4.2.3 Physician access | 0.18 |
| **Evacuation** | |
| 4.2.4. Access and evacuation | 0.73 |
| 4.2.5 Transportation | 0.82 |
| 4.2.6 Evacuation routes | 0.36 |
| **Supplies** | |
| 4.2.7 Efficient energy use | 1 |
| 4.2.8 Efficient water use | 0.64 |
| 4.2.9 Gas | 0.45 |
| 4.2.10 Industrial resupply potential | 0.27 |
| 4.2.11 Wastewater treatment | 0.55 |
| **Communication** | |
| 4.2.12 Telecommunication | 0.73 |
| 4.2.13 High-speed Internet infrastructure | 0.18 |

Step 3: Inference – Fuzzy logic

The design of the hierarchical framework and the calculation of the weighting factors of the variables within the analyzed dimension allow implementing fuzzy logic as an inference system. That is, weighting factors are useful to determine fuzzy rules for aggregating indicators and components. The aggregation is done by following the relationships between the variables provided by the hierarchical model.

The following steps to implement the fuzzy logic inference system are performed:

*Step 1: Fuzzification process – membership functions*

As mentioned before, a set of parameters is used to define the functionality of PEOPLES indicators. The proposed methodology adopts three of the four functionality parameters, namely initial functionality $q_0$, drop of functionality, which is defined as $\Delta q = q_0 - q_1$, where $q_1$ is the functionality after the event, and the restoration time $T_r$. These parameters could have different states called linguistic quantifiers or fuzzy sets. To implement the fuzzy inference system in the PEOPLES framework easily, the number of states is set to three states for all indicators' parameters: *low, medium*, and *high* for the functionality parameters, *short, long,* and *very long* for the recovery time parameter, and *resilient*, *intermediate*, and *not resilient* for the resilience index. Considering more than 3 states leads to a more complicated fuzzy process. That is, if more states are considered, more membership functions would then be necessary, and a high number of fuzzy rules would be required to cover all the possible permutations of the states. The membership functions considered in the methodology are based on trapezoidal fuzzy numbers and they are expressed by four vertices (a, b, c, and d). For instance, considering the functionality after the disaster ($q_1$) of the *Telecommunication* indicator, its membership degree ($\mu$) to the $i^{th}$ granularity ($A_i$) will be:

$$\mu_{A_i}(q_1) = \begin{cases} 0, & 0 \leq q_1 \leq a \\ \frac{q_1 - a}{b - a}, & a < q_1 < b \\ 1, & b \leq q_1 \leq c \\ \frac{q_1 - d}{c - d}, & c < q_1 < d \\ 0, & d \leq q_1 \end{cases} \quad (2)$$

The membership functions have been first designed relying on authors opinion and experience, then they have been calibrated so that the resilience outcome predicted by the model is equal (or nearly equal) to the resilience outcome obtained in the benchmark case study [19]. Calibration is a fundamental operation and consists of gradually modifying the shapes of the membership functions such that the final output and approximately matches that of the benchmark case study. An example of granulation assigned to the initial functionality $q_0$ of the *Telecommunication* indicator and *Resilience* indicator is illustrated in Figure 13.

Figure 13. Membership function and granulation for the initial functionality $q_0$ of *Telecommunication* indicator and *Resilience* indicator

The functionality and recovery time parameters for each indicator and component listed in Table 1 are used as numerical inputs in the fuzzification process. In the case of numerical values used as inputs, transformation values on a range [0 1] are not required. That is, one can enter the corresponding membership graph using directly the numerical values listed in Table 1 and obtain the membership degree. The membership functions used in the methodology associated with the *Physical Infrastructure* dimension along with its components and indicators are based on trapezoidal fuzzy numbers (see Figure 13) and they are listed in Table 4.

Table 4. Membership functions for Physical Infrastructure dimension, components, and indicators

| Dimension/component/ subgroups/indicators | Initial functionality $q_0$ ($\mu_L, \mu_M, \mu_H$) | Drop of functionality $\Delta_q$ ($\mu_L, \mu_M, \mu_H$) | Repair time $T_r$ ($\mu_S, \mu_L, \mu_{VL}$) |
|---|---|---|---|
| 4 - *Physical infrastructure* | (0, 0, 0.1, 0.3), (0.2, 0.3, 0.6, 0.8), (0.5, 0.9, 1, 1) | (0, 0, 0.3, 0.5), (0.2, 0.5, 0.7, 0.8), (0.7, 0.9, 1, 1) | (0, 0, 0.1, 0.2), (0.1, 0.3, 0.7, 0.8), (0.7, 0.9, 1, 1) |
| 4-1 - *Facilities* | (0, 0, 0.1, 0.4), (0.1, 0.3, 0.6, 0.9), (0.6, 0.9, 1, 1) | (0, 0, 0.3, 0.5), (0.2, 0.5, 0.7, 0.8), (0.7, 0.9, 1, 1) | (0, 0, 0.1, 0.4), (0.2, 0.3, 0.7, 0.8), (0.7, 0.9, 1, 1) |
| **Housing** | (0, 0, 0.1, 0.4), (0.1, 0.3, 0.6, 0.9), (0.6, 0.9, 1, 1) | (0, 0, 0.3, 0.5), (0.2, 0.5, 0.7, 0.8), (0.7, 0.9, 1, 1) | (0, 0, 0.1, 0.2), (0.1, 0.3, 0.7, 0.8), (0.7, 0.9, 1, 1) |
| 4-1-1 - Sturdier housing types | (0, 0, 0.1, 0.4), (0.1, 0.3, 0.6, 0.9), (0.6, 0.9, 1, 1) | (0, 0, 0.3, 0.5), (0.3, 0.5, 0.7, 0.8), (0.7, 0.9, 1, 1) | [0, 0, 0.1, 0.2], [0.1, 0.3, 0.7, 0.8], [0.7, 0.9, 1, 1] |
| 4-1-2 - Temporary housing availability | (0, 0, 0.1, 0.4), (0.1, 0.3, 0.6, 0.9), (0.6, 0.9, 1, 1) | (0, 0, 0.3, 0.5), (0.3, 0.5, 0.7, 0.8), (0.7, 0.9, 1, 1) | [0, 0, 0.1, 0.2], [0.1, 0.3, 0.7, 0.8], [0.7, 0.9, 1, 1] |
| 4-1-3 - Housing stock construction quality | (0, 0, 0.1, 0.4), (0.1, 0.3, 0.6, 0.9), (0.6, 0.9, 1, 1) | (0, 0, 0.3, 0.5), (0.3, 0.5, 0.7, 0.8), (0.7, 0.9, 1, 1) | [0, 0, 0.1, 0.2], [0.1, 0.3, 0.7, 0.8], [0.7, 0.9, 1, 1] |
| **Commercial Activities** | (0, 0, 0.1, 0.4), (0.1, 0.3, 0.6, 0.9), (0.6, 0.9, 1, 1) | (0, 0, 0.3, 0.5), (0.3, 0.5, 0.7, 0.8), (0.7, 0.9, 1, 1) | [0, 0, 0.1, 0.2], [0.1, 0.3, 0.7, 0.8], [0.7, 0.9, 1, 1] |
| 4-1-4 - Economic infrastructure exposure | (0, 0, 0.1, 0.4), (0.1, 0.3, 0.6, 0.9), (0.6, 0.9, 1, 1) | (0, 0, 0.3, 0.5), (0.3, 0.5, 0.7, 0.8), (0.7, 0.9, 1, 1) | [0, 0, 0.1, 0.2], [0.1, 0.3, 0.7, 0.8], [0.7, 0.9, 1, 1] |
| 4-1-5 - Distribution commercial facilities | (0, 0, 0.1, 0.4), (0.1, 0.3, 0.6, 0.9), (0.6, 0.9, 1, 1) | (0, 0, 0.3, 0.5), (0.3, 0.5, 0.7, 0.8), (0.7, 0.9, 1, 1) | [0, 0, 0.1, 0.2], [0.1, 0.3, 0.7, 0.8], [0.7, 0.9, 1, 1] |
| **Services** | (0, 0, 0.1, 0.4), (0.1, 0.3, 0.6, 0.9), (0.6, 0.9, 1, 1) | (0, 0, 0.3, 0.5), (0.3, 0.5, 0.7, 0.8), (0.7, 0.9, 1, 1) | [0, 0, 0, 0.3], [0.1, 0.3, 0.7, 0.8], [0.7, 0.9, 1, 1] |
| 4-1-6 - Community services | (0, 0, 0.1, 0.4), (0.1, 0.3, 0.6, 0.9), (0.6, 0.9, 1, 1) | (0, 0, 0.3, 0.5), (0.3, 0.5, 0.7, 0.8), (0.7, 0.9, 1, 1) | [0, 0, 0.1, 0.2], [0.1, 0.3, 0.7, 0.8], [0.7, 0.9, 1, 1] |
| 4-1-7 - Hotels and accommodations | (0, 0, 0.1, 0.4), (0.1, 0.3, 0.6, 0.9), (0.6, 0.9, 1, 1) | (0, 0, 0.3, 0.5), (0.3, 0.5, 0.7, 0.8), (0.7, 0.9, 1, 1) | [0, 0, 0, 0.3], [0.1, 0.3, 0.7, 0.8], [0.7, 0.9, 1, 1] |
| 4-1-8 - Schools | (0, 0, 0.1, 0.4), (0.1, 0.3, 0.6, 0.9), (0.6, 0.9, 1, 1) | (0, 0, 0.3, 0.5), (0.3, 0.5, 0.7, 0.8), (0.7, 0.9, 1, 1) | [0, 0, 0, 0.3], [0.1, 0.3, 0.7, 0.8], [0.7, 0.9, 1, 1] |
| 4-2 - *Lifelines* | (0, 0, 0.1, 0.3), (0.2, 0.4, 0.6, 0.8), (0.6, 0.9, 1, 1) | (0, 0, 0.3, 0.5), (0.2, 0.5, 0.7, 0.8), (0.7, 0.9, 1, 1) | [0, 0, 0.1, 0.3], [0.1, 0.3, 0.7, 0.8], [0.7, 0.9, 1, 1] |
| **Healthcare** | (0, 0, 0.1, 0.4), (0.1, 0.3, 0.6, 0.9), (0.6, 0.9, 1, 1) | (0, 0, 0.3, 0.5), (0.3, 0.5, 0.7, 0.8), (0.7, 0.9, 1, 1) | [0, 0, 0.1, 0.2], [0.1, 0.2, 0.7, 0.8], [0.7, 0.9, 1, 1] |
| 4-2-1 - Mental health support | (0, 0, 0.1, 0.4), (0.1, 0.3, 0.6, 0.9), (0.6, 0.9, 1, 1) | (0, 0, 0.3, 0.5), (0.3, 0.5, 0.7, 0.8), (0.7, 0.9, 1, 1) | [0, 0, 0.1, 0.2], [0.1, 0.2, 0.7, 0.8], [0.7, 0.9, 1, 1] |
| 4-2-2 - Medical care capacity | (0, 0, 0.1, 0.4), (0.1, 0.3, 0.6, 0.9), (0.6, 0.9, 1, 1) | (0, 0, 0.3, 0.5), (0.3, 0.5, 0.7, 0.8), (0.7, 0.9, 1, 1) | [0, 0, 0.1, 0.2], [0.1, 0.2, 0.7, 0.8], [0.7, 0.9, 1, 1] |
| 4-2-3 - Physician access | (0, 0, 0.1, 0.4), (0.1, 0.3, 0.6, 0.9), (0.6, 0.9, 1, 1) | (0, 0, 0.3, 0.5), (0.3, 0.5, 0.7, 0.8), (0.7, 0.9, 1, 1) | [0, 0, 0.1, 0.2], [0.1, 0.2, 0.7, 0.8], [0.7, 0.9, 1, 1] |
| **Evacuation** | (0, 0, 0.1, 0.4), (0.1, 0.3, 0.6, 0.9), (0.6, 0.9, 1, 1) | (0, 0, 0.3, 0.5), (0.3, 0.5, 0.7, 0.8), (0.7, 0.9, 1, 1) | [0, 0, 0.1, 0.2], [0.1, 0.2, 0.7, 0.8], [0.7, 0.9, 1, 1] |
| 4-2-4 - Access and evacuation | (0, 0, 0.1, 0.4), (0.1, 0.3, 0.6, 0.9), (0.6, 0.9, 1, 1) | (0, 0, 0.3, 0.5), (0.3, 0.5, 0.7, 0.8), (0.7, 0.9, 1, 1) | (0, 0, 0.1, 0.2], [0.1, 0.2, 0.7, 0.8], [0.7, 0.9, 1, 1] |
| 4-2-5 - Transportation | (0, 0, 0.1, 0.4), (0.1, 0.3, 0.6, 0.9), (0.6, 0.9, 1, 1) | (0, 0, 0.3, 0.5), (0.3, 0.5, 0.7, 0.8), (0.7, 0.9, 1, 1) | [0, 0, 0.1, 0.2], [0.1, 0.2, 0.7, 0.8], [0.7, 0.9, 1, 1] |
| 4-2-6 - Evacuation routes | (0, 0, 0.1, 0.4), (0.1, 0.3, 0.6, 0.9), (0.6, 0.9, 1, 1) | (0, 0, 0.3, 0.5), (0.3, 0.5, 0.7, 0.8), (0.7, 0.9, 1, 1) | [0, 0, 0.1, 0.2], [0.1, 0.2, 0.7, 0.8], [0.7, 0.9, 1, 1] |
| **Supplies** | (0, 0, 0.1, 0.4), (0.1, 0.3, 0.6, 0.9), (0.6, 0.9, 1, 1) | [0, 0, 0.3, 0.4], [0.2, 0.3, 0.7, 0.8], [0.7, 0.9, 1, 1] | [0, 0, 0.1, 0.2], [0.1, 0.3, 0.7, 0.8], [0.7, 0.9, 1, 1] |
| 4-2-7 - Efficient energy use | (0, 0, 0.1, 0.4), (0.1, 0.3, 0.6, 0.9), (0.6, 0.9, 1, 1) | [0, 0, 0.3, 0.5], [0.3, 0.5, 0.7, 0.8], [0.7, 0.9, 1, 1] | [0, 0, 0.1, 0.1], [0.1, 0.3, 0.7, 0.8], [0.7, 0.9, 1, 1] |
| 4-2-8 - Efficient Water Use | (0, 0, 0.1, 0.4), (0.1, 0.3, 0.6, 0.9), (0.6, 0.9, 1, 1) | [0, 0, 0.3, 0.5], [0.3, 0.5, 0.7, 0.8], [0.7, 0.9, 1, 1] | [0, 0, 0.1, 0.1], [0.1, 0.3, 0.7, 0.8], [0.7, 0.9, 1, 1] |
| 4-2-9 - Gas | (0, 0, 0.1, 0.4), (0.1, 0.3, 0.6, 0.9), (0.6, 0.9, 1, 1) | [0, 0, 0.3, 0.5], [0.3, 0.5, 0.7, 0.8], [0.7, 0.9, 1, 1] | [0, 0, 0.1, 0.1], [0.1, 0.3, 0.7, 0.8], [0.7, 0.9, 1, 1] |
| 4-2-10 - Industrial re-supply potential | (0, 0, 0.1, 0.4), (0.1, 0.3, 0.6, 0.9), (0.6, 0.9, 1, 1) | [0, 0, 0.3, 0.5], [0.2, 0.3, 0.7, 0.8], [0.7, 0.9, 1, 1] | [0, 0, 0.1, 0.2], [0.1, 0.3, 0.7, 0.8], [0.7, 0.9, 1, 1] |
| 4-2-11 - Waste water treatment | (0, 0, 0.1, 0.4), (0.1, 0.3, 0.6, 0.9), (0.6, 0.9, 1, 1) | [0, 0, 0.3, 0.5], [0.2, 0.3, 0.7, 0.8], [0.7, 0.9, 1, 1] | [0, 0, 0.1, 0.2], [0.1, 0.3, 0.7, 0.8], [0.7, 0.9, 1, 1] |
| **Communication** | (0, 0, 0.1, 0.4), (0.1, 0.3, 0.6, 0.9), (0.6, 0.9, 1, 1) | [0, 0, 0.3, 0.5], [0.2, 0.4, 0.7, 0.8], [0.7, 0.9, 1, 1] | [0, 0, 0.3, 0.4], [0.2, 0.4, 0.7, 0.8], [0.7, 0.9, 1, 1] |

| 4-2-12 - Telecommunication | (0, 0, 0.1, 0.4), (0.1, 0.3, 0.6, 0.9), (0.6, 0.9, 1, 1) | [0, 0, 0.3, 0.5], [0.2, 0.4, 0.7, 0.8], [0.7, 0.9, 1, 1] | [0, 0, 0.3, 0.4], [0.2, 0.4, 0.7, 0.8], [0.7, 0.9, 1, 1] |
| --- | --- | --- | --- |
| 4-2-13 - High-speed internet infrastructure | (0, 0, 0.1, 0.4), (0.1, 0.3, 0.6, 0.9), (0.6, 0.9, 1, 1) | [0, 0, 0.3, 0.5], [0.2, 0.4, 0.7, 0.8], [0.7, 0.9, 1, 1] | [0, 0, 0.3, 0.4], [0.2, 0.4, 0.7, 0.8], [0.7, 0.9, 1, 1] |

The membership degrees obtained through the fuzzification process for the components under the *Physical Infrastructure* dimension are listed in Table 5.

Table 5. Fuzzification process

| Dimension/component/ subgroups/indicators | Fuzzification | | |
| --- | --- | --- | --- |
| | Initial functionality $q_0$ ($\mu_L, \mu_M, \mu_H$) | Drop of functionality $\Delta_q$ ($\mu_L, \mu_M, \mu_H$) | Repair time $T_r$ ($\mu_S, \mu_L, \mu_{VL}$) |
| 4 - *Physical infrastructure* | (0,0,0.63) | (0.92, 0.37,0) | (0,0.39,0) |
| 4-1 - *Facilities* | (0,0.67,0.33) | (0.37,0.63,0) | (0.31,0.69,0) |
| **Housing** | (0,1,0) | (0.62,0.38,0) | (0,1,0) |
| **Commercial Activities** | (0,0.53,0.47) | (1,0,0) | (0.95,0,0) |
| **Services** | (0,0.76,0.24) | (1,0,0) | (0.23,0.77,0) |
| 4-2 - *Lifelines* | (0,0,0.76) | (0.92, 0.37,0) | (0.9, 0.1, 0) |
| **Healthcare** | (0,0.78,0.22) | (1,0,0) | (0.72,0,0) |
| **Evacuation** | (0,0.15,0.85) | (1,0,0) | (1,0,0) |
| **Supplies** | (0.36,0.64,0) | (0,1,0) | (0,1,0) |
| **Communication** | (0,0.15,0.85) | (0.64,0.82,0) | (0.93, 0.52,0) |

*Step 2: Aggregation through Fuzzy rules*

The most common type of *fuzzy rule base* (FRB) known as the Mamdani type, is adopted herein. It consists of three connectives: the aggregation of the antecedents in each fuzzy rule (AND connectives), implication (IF-THEN connectives), and aggregation of the rules (ALSO connectives).

As shown in Figure 7, many indicators with their corresponding parameters are considered in the physical infrastructure framework, and consequently, several fuzzy rules are necessary to combine them. In a fuzzy-based model, a high number of inputs results in an exponential increase in the number of fuzzy rules. Different techniques have been developed in the literature to cope with the high number of rules, such as identification of functional relationships, interpolation, and rule hierarchy. In this paper, a decomposition technique at the level of indicators proposed by [47], is adopted. The hierarchical structure can be decomposed at the level of indicators by introducing intermediate connections through intermediate rules among the indicators at different levels of the framework resulting in a maximum of $3^3 = 27$ rules per subgroup. As shown, no more than three indicators in each subgroup are aggregated through intermediate rules (temporary rules), which are for example $TR_1$, $TR_2$, $TR_3$, etc. The output of the intermediate inference is combined through fuzzy rule based $R_1$ and $R_2$. For instance, indicators within the subgroup *Services* are aggregated through $TR_1$, indicators under the subgroup *Commercial Activities* are combined through $TR_2$, and finally, the indicators under the subgroup *Housing* are aggregated through $TR_3$. The outputs of these components are then aggregated through $R_1$ to obtain the *Facilities* component. At each level of the hierarchical scheme, the three-tuple fuzzy set output is defuzzified to obtain a single crisp value. In turn, this value is fuzzified into the next level. In this work, fuzzy rules have been determined accordingly to the weighting factors calculated in step 2. That is, assuming to mapping the three granularities [L, M, H] into the numerical values [$F_L$, $F_M$, $F_H$] = [1,2,3], which indicate an increase of functionality ($F$) of the system, and considering two inputs with $w_1 = 0.75$ and $w_2 = 0.5$ respectively, the following empirical statements are used to evaluate rules:

$$\begin{aligned}
&IF\ (1.25 \leq \sum_{i=1}^{n} F_i * w_i \leq 1.56)\ THEN\ (output\ L) \\
&IF\ (1.56 \leq \sum_{i=1}^{n} F_i * w_i \leq 1.87)\ AND\ (1.87 \leq \sum_{i=1}^{n} F_i * w_i \leq 2.81)\ THEN\ (output\ is\ M) \\
&IF\ (2.81 < \sum_{i=1}^{n} F_i * w_i \leq 3.75)\ THEN\ (output\ is\ H)
\end{aligned} \quad (3)$$

where $w_i$ is the weighting factor for indicator $i$, $F_i$ is the numerical value corresponding to the belonging granularity of indicator $i$, and $n$ is the total number of indicators in the subgroups. The thresholds in Eq. (3) are empirically selected based on the weighting factors associated with each variable and they are based on the minimum and maximum value of the summations. Therefore, they change in every aggregation.

An example of the fuzzy rules assigned for combining the recovery time parameter of the *Commercial Activities* indicators is given in Table 6.

Table 6. Fuzzy rule table for $T_r$ of *Commercial Activities* indicator

| Rule | Economic infrastructure exposure<br>w = 0.75 | Distribution commercial facilities<br>w = 0.5 | Summation | Commercial Activities |
|---|---|---|---|---|
| 1 | S | S | 1.25 | S |
| 2 | S | L | 1.75 | L |
| 3 | S | VL | 2.25 | L |
| 4 | L | S | 2 | L |
| 5 | L | L | 2.5 | L |
| 6 | L | VL | 3 | VL |
| 7 | VL | S | 2.75 | L |
| 8 | VL | L | 3.25 | VL |
| 9 | VL | VL | 3.75 | VL |

Table 6 shows that the output is mainly driven by the *Economic infrastructure exposure* indicator (w = 0.75), in agreement with the fact that it is more important than the *Distribution commercial facilities* indicator (w = 0.5).

Using the fuzzy rule table (Table 6), the recovery time $T_R$ parameter of the *Commercial Activities* indicator is computed as follows:

$$\begin{aligned} \mu_S^{CA} &= \max\,(\min(1,0.95)\,,\min(1,0),\min\,(0,0.95) = 0.95 \\ \mu_L^{CA} &= \max\,(\min(1,0)\,,\min(0,0.\,)\,,\min(0,0.95)) = 0 \\ \mu_{VL}^{CA} &= \max(\min(0,0)\,,\min(0,0)\,,\min\,(0,0) = 0 \end{aligned} \quad (4)$$

*Step 3: Defuzzification process – crisp output*

The last step of the fuzzy-based methodology is the defuzzification process. Here the center of gravity (also called the center of area) method is used. The method first calculates the area under the membership functions and within the range of the output variable, then calculates the geometric center of the area as follows:

$$CoA = \frac{\int_{x_{min}}^{x_{max}} f(x) \cdot x \, dx}{\int_{x_{min}}^{x_{max}} f(x) \, dx} \quad (5)$$

where *f(x)* is the function that shapes the output fuzzy set after the aggregation process and *x* stands for the real values inside the fuzzy set support [0,1].

Using the center of gravity method, the recovery time parameter $T_R$ of the *Commercial Activities* indicator is defuzzified as 0.086. The defuzzification of the other indicators and components is done similarly. The resilience index of the *Physical Infrastructure* dimension is given through inferencing the *Physical Infrastructure* functionality and recovery time parameters. The results obtained in terms of fuzzy functionalities and recovery time are listed in Table 7.

Table 7. Fuzzy functionality and recovery time parameters for *Lifelines* and *Facilities* components and the *Physical Infrastructure* dimension

| | Parameters | Results |
|---|---|---|
| *Lifelines* component | $q_0$ | 0.67 |
| | $\Delta q$ | 0.328 |
| | $q_1$ | 0.342 |
| | $Tr$ | 0.117 |
| *Facilities* component | $q_0$ | 0.67 |
| | $\Delta q$ | 0.328 |
| | $q_1$ | 0.342 |
| | $Tr$ | 0.329 |
| *Physical Infrastructure* dimension | $q_0$ | 0.80 |
| | $\Delta q$ | 0.31 |
| | $q_1$ | 0.49 |
| | $Tr$ | 0.2 |

The resilience index $R$ of the city of San Francisco is computed as $R = 0.73$. The $R$ is a percentage that reflects the response of the community to the earthquake event. That is, a higher $R$ signifies a good response of the community. In this demonstrative example, the obtained value of $R$ corresponds only to the physical infrastructure dimension of the community. To establish a resilience index for a whole community, the functionality and recovery time parameters of other dimensions have to be similarly evaluated and combined in the same way the available measures were aggregated.

The loss of resilience of the *Physical Infrastructure* dimension can be computed using the following equation:

$$LOR_{PhysicalInfrastructure} = 1 - R_{PhysicalInfrastructure} = 27\% \qquad (6)$$

Finally, the functionality curves for the *Lifelines* and *Facilities* components and for the *Physical Infrastructure dimension* are shown in Figure 12.

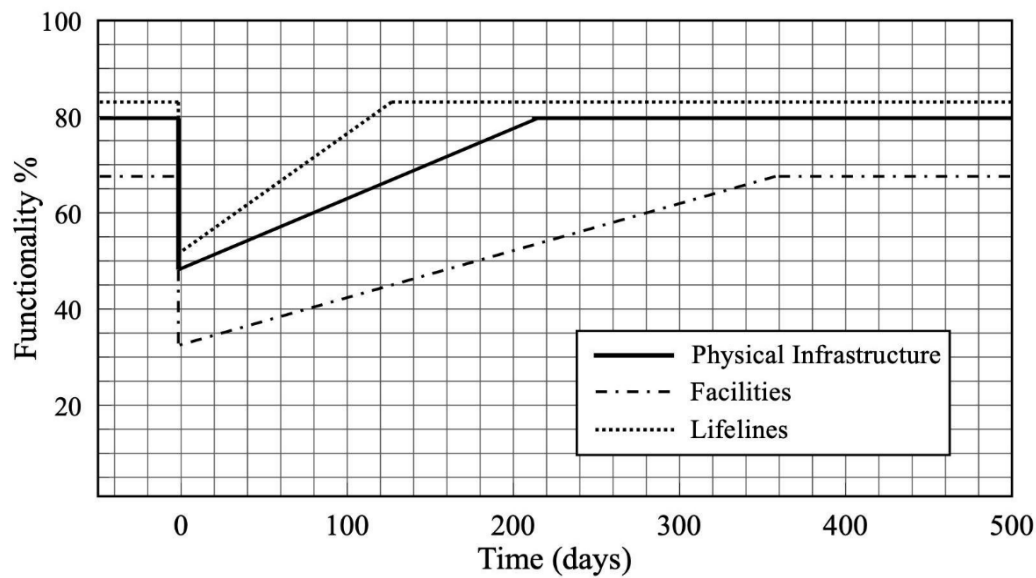

Figure 14. Functionality curves of components *Facilities* and *Lifelines* under the dimension *Physical Infrastructure*

From Figure 14 it is possible comparing the functionality curves of the two components facilities and lifelines. The city of San Francisco shows more problems in facilities than lifelines. That is, it is evident that the *LOR* of facilities is higher than lifelines. In such a case, authorities should focus more on improving facilities to get great benefits.

The proposed methodology has been verified by comparing the obtained *R* with the result given by [19], who analyzed the same case study focusing on the estimation of the loss of resilience *LOR*. The verification phase has been conducted at each level of the framework by calibrating the shape of membership functions that have a strong impact on final results. Within the proposed approach, the shape of membership functions was first estimated through the authors' opinion and it was designed to be as symmetrical as possible; then the angle points of the membership functions were modified little by little to get *R*, and consequently *LOR*, as similar as possible to the result obtained from the benchmark system [19] . Thus, after the calibration, the membership functions used in the methodology are not equivalent (e.g., the wideness of the membership function "low" may be larger than "high").

It should be noted that looking at a single resilience index sometimes leads to losing information about indicators that suffer from resilience deficiencies and need to be improved. To manage or improve resilience, close attention should be paid to the individual indicators that influence system resilience to highlight the strengths and weaknesses of system resilience.

# 6 Conclusion

The recent disasters worldwide have demonstrated that resilience is the solution to cope with natural and man-made threats. This paper presents a holistic framework for evaluating community resilience in response to a catastrophic event. The proposed methodology benefits from the structure of the PEOPLES framework for its implementation and deals with the complexity and vagueness that characterizes processes where human intervention is significant by implementing fuzzy logic theory.

A general indicator-based resilience model to estimate the resilience index of communities is proposed. The resilience assessment can be easily adapted to any communities of different sizes and types by changing the values of the indicators. The fuzzy logic inference method is introduced within the resilience assessment model to deal with the uncertainties that may affect the indicator collection. That is, the model considers uncertainties associated with each indicator as it does not require deterministic and precise input data. Since indicators do not contribute equally to the resilience assessment, the contribution of every indicator towards resilience has been determined through a proposed interdependency analysis, which resulted in an importance factor assigned to each indicator.

To illustrate the applicability of the resilience assessment model, a case study of an earthquake that struck the city of San Francisco on October 17, 1989, with a magnitude of 6.0 on the Richter scale is considered as a hazard event. Such a case study was used to verify the proposed methodology by comparing the model output with the result of [19], who proposed a methodology to treat the PEOPLES framework as a quantitative model by analyzing the same case study. The verification of the fuzzy-based resilience assessment was performed by calibrating the membership functions of fuzzy sets to obtain a resilience index that approximates that of the benchmark system. Therefore, the obtained resilience index depends on the decisions made during the design of the fuzzy inference system. The results have shown that the proposed approach is consistent with the existing evaluations in the literature and can be easily applied to large communities. The results from the proposed framework are suitable for assisting decision-makers, planners, and engineers to assess and learn from the resilience of their communities in the face of a particular event. It is worth noting that paying attention to a single resilience index alone sometimes causes system managers to lose information about the indicators that suffer from resilience deficiencies and thereby need to be improved. Therefore, paying attention to the factors influencing system resilience can bring out the strengths and weaknesses of system resilience in either the technical or organizational domain. It helps managers to identify the area of the system that are more sensitive and require more attention. Consequently, they can make a more detailed assessment and implement improvement strategies.

Future work will focus on applying the proposed approach by considering all dimensions of PEOPLES as more reliable data become available.

**CrediT authorship contribution statement**

**Melissa De Iuliis:** Writing – original draft, Methodology, Software, Verification. **Omar Kammouh:** Conceptualization, Supervision, Writing – review & editing. **Gian Paolo Cimellaro:** Supervision, Writing – review & editing, Funding acquisition.

**Acknowledgment**

The research leading to these results has received funding from the European Research Council under the Grant Agreement n° 637842 of the project IDEAL RESCUE Integrated Design and Control of Sustainable Communities during Emergencies

# References

[1]  Allenby, B. and J. Fink, *Toward inherently secure and resilient societies.* Science, 2005. **309**(5737): p. 1034-1036.

[2]  Bruneau, M., et al., *A framework to quantitatively assess and enhance the seismic resilience of communities.* Earthquake spectra, 2003. **19**(4): p. 733-752.

[3]  Cimellaro, G.P., A.M. Reinhorn, and M. Bruneau, *Framework for analytical quantification of disaster resilience.* Engineering structures, 2010. **32**(11): p. 3639-3649.

[4]  Renschler, C.S., et al. *Developing the 'PEOPLES' resilience framework for defining and measuring disaster resilience at the community scale.* in *Proceedings of the 9th US national and 10th Canadian conference on earthquake engineering.* 2010.

[5]  Cimellaro, G.P., *Urban resilience for emergency response and recovery.* Switzerland: Springer [DOI: 10.1007/978-3-319-30656-8], 2016.

[6]  Chang, S.E. and M. Shinozuka, *Measuring improvements in the disaster resilience of communities.* Earthquake spectra, 2004. **20**(3): p. 739-755.

[7]  Gilbert, S. and B.M. Ayyub, *Models for the Economics of Resilience.* ASCE-ASME Journal of Risk and Uncertainty in Engineering Systems, Part A: Civil Engineering, 2016. **2**(4): p. 04016003.

[8]  Liu, X., E. Ferrario, and E. Zio, *Resilience analysis framework for interconnected critical infrastructures.* ASCE-ASME Journal of Risk and Uncertainty in Engineering Systems, Part B: Mechanical Engineering, 2017. **3**(2).

[9]  Ayyub, B.M., *Practical resilience metrics for planning, design, and decision making.* ASCE-ASME Journal of Risk and Uncertainty in Engineering Systems, Part A: Civil Engineering, 2015. **1**(3): p. 04015008.

[10] Scherzer, S., P. Lujala, and J.K. Rød, *A community resilience index for Norway: An adaptation of the Baseline Resilience Indicators for Communities (BRIC).* International Journal of Disaster Risk Reduction, 2019. **36**: p. 101107.

[11] UNISDR, U. *Hyogo framework for action 2005–2015: Building the resilience of nations and communities to disasters.* in *Extract from the final report of the World Conference on Disaster Reduction (A/CONF. 206/6).* 2005. The United Nations International Strategy for Disaster Reduction Geneva.

[12] UNISDR, U. *Sendai framework for disaster risk reduction 2015–2030.* in *Proceedings of the 3rd United Nations World Conference on DRR, Sendai, Japan.* 2015.

[13] Kammouh, O., G. Dervishaj, and G.P. Cimellaro, *Quantitative framework to assess resilience and risk at the country level.* ASCE-ASME Journal of risk and uncertainty in engineering systems, part A: civil engineering, 2018. **4**(1): p. 04017033.

[14] Cutter, S.L., K.D. Ash, and C.T. Emrich, *The geographies of community disaster resilience.* Global environmental change, 2014. **29**: p. 65-77.

[15] SPUR, S.F.P.a.U.R.A., *Defining what San Francisco needs from its seismic mitigation policies.* 2009.

[16] Kwasinski, A., et al., *A conceptual framework for assessing resilience at the community scale.* Gaithersburg, MD: National Institute of Standards and Technology, 2016: p. 16-001.

[17] Cimellaro, G.P., et al., *PEOPLES: a framework for evaluating resilience.* Journal of Structural Engineering, 2016. **142**(10): p. 04016063.

[18] Kammouh, O., et al., *Deterministic and fuzzy-based methods to evaluate community resilience.* Earthquake Engineering and Engineering Vibration, 2018. **17**(2): p. 261-275.

[19] Kammouh, O., et al., *Resilience assessment of urban communities.* ASCE-ASME Journal of Risk and Uncertainty in Engineering Systems, Part A: Civil Engineering, 2019. **5**(1): p. 04019002.

[20] Cohen, O., et al., *Building resilience: The relationship between information provided by municipal authorities during emergency situations and community resilience.* Technological Forecasting and Social Change, 2017. **121**: p. 119-125.

[21] Pfefferbaum, B., R.L. Pfefferbaum, and R.L. Van Horn, *Community resilience interventions: Participatory, assessment-based, action-oriented processes.* American Behavioral Scientist, 2015. **59**(2): p. 238-253.

[22] White, R.K., et al., *A practical approach to building resilience in America's communities.* American Behavioral Scientist, 2015. **59**(2): p. 200-219.

[23]   Schultz, M.T. and E.R. Smith, *Assessing the resilience of coastal systems: A probabilistic approach.* Journal of Coastal Research, 2016. **32**(5): p. 1032-1050.
[24]   Kammouh, O., P. Gardoni, and G.P. Cimellaro, *Probabilistic framework to evaluate the resilience of engineering systems using Bayesian and dynamic Bayesian networks.* Reliability Engineering & System Safety, 2020. **198**: p. 106813.
[25]   Franchin, P. and F. Cavalieri, *Probabilistic assessment of civil infrastructure resilience to earthquakes.* Computer-Aided Civil and Infrastructure Engineering, 2015. **30**(7): p. 583-600.
[26]   Cai, H., et al., *Modeling the dynamics of community resilience to coastal hazards using a Bayesian network.* Annals of the American Association of Geographers, 2018. **108**(5): p. 1260-1279.
[27]   Kameshwar, S., et al., *Probabilistic decision-support framework for community resilience: Incorporating multi-hazards, infrastructure interdependencies, and resilience goals in a Bayesian network.* Reliability Engineering & System Safety, 2019. **191**: p. 106568.
[28]   Baraldi, P., et al., *Comparing the treatment of uncertainty in Bayesian networks and fuzzy expert systems used for a human reliability analysis application.* Reliability Engineering & System Safety, 2015. **138**: p. 176-193.
[29]   Renschler, C., et al., *The PEOPLES Resilience Framework: A conceptual approach to quantify community resilience.* Proceedings of COMPDYN, 2011: p. 26-28.
[30]   Cutter, S.L., C.G. Burton, and C.T. Emrich, *Disaster resilience indicators for benchmarking baseline conditions.* Journal of homeland security and emergency management, 2010. **7**(1).
[31]   Kammouh, O., et al. *PEOPLES: Indicator-Based Tool To Compute Community Resilience*. in *Eleventh US National Conference on Earthquake Engineering Integrating Science, Engineering & Policy June*. 2018.
[32]   Zadeh, L.A., *Fuzzy sets.* Information and control, 1965. **8**(3): p. 338-353.
[33]   Kabir, G. and R.S. Sumi, *Power substation location selection using fuzzy analytic hierarchy process and PROMETHEE: A case study from Bangladesh.* Energy, 2014. **72**: p. 717-730.
[34]   Tesfamariam, S., R. Sadiq, and H. Najjaran, *Decision making under uncertainty—An example for seismic risk management.* Risk Analysis: An International Journal, 2010. **30**(1): p. 78-94.
[35]   De Iuliis, M., et al., *Downtime estimation of building structures using fuzzy logic.* International journal of disaster risk reduction, 2019. **34**: p. 196-208.
[36]   Quelch, J. and I.T. Cameron, *Uncertainty representation and propagation in quantified risk assessment using fuzzy sets.* Journal of loss prevention in the process industries, 1994. **7**(6): p. 463-473.
[37]   Bonvicini, S., P. Leonelli, and G. Spadoni, *Risk analysis of hazardous materials transportation: evaluating uncertainty by means of fuzzy logic.* Journal of Hazardous Materials, 1998. **62**(1): p. 59-74.
[38]   Tesfamariam, S. and M. Saatcioglu, *Seismic vulnerability assessment of reinforced concrete buildings using hierarchical fuzzy rule base modeling.* Earthquake Spectra, 2010. **26**(1): p. 235-256.
[39]   Colangelo, F., *Probabilistic characterisation of an analytical fuzzy-random model for seismic fragility computation.* Structural safety, 2013. **40**: p. 68-77.
[40]   Karimi, I. and K. Meskouris, *Risk management of natural disasters: a fuzzy-probabilistic methodology and its application to seismic hazard.* 2006, Fakultät für Bauingenieurwesen.
[41]   Tesfamariam, S. and M. Saatcioglu, *Risk-based seismic evaluation of reinforced concrete buildings.* Earthquake Spectra, 2008. **24**(3): p. 795-821.
[42]   UNISDR, *How to make cities more resilient: A handbook for local government leaders: A contribution to the global campaign 2010-2015: Making cities resilient - My city is getting ready!* 2012.
[43]   Norris, F.H., et al., *Community resilience as a metaphor, theory, set of capacities, and strategy for disaster readiness.* American journal of community psychology, 2008. **41**(1): p. 127-150.
[44]   George, K. and Y. Bo, *Fuzzy sets and fuzzy logic, theory and applications.* 2008.
[45]   Cimellaro, G.P., et al., *Rapid building damage assessment system using mobile phone technology.* Earthquake Engineering and Engineering Vibration, 2014. **13**(3): p. 519-533.
[46]   US Census Bureau, *Selected housing characteristics, 2007-2011 American Community Survey 5-year estimates.* 2010.
[47]   Magdalena, L., *On the role of context in hierarchical fuzzy controllers.* International journal of intelligent systems, 2002. **17**(5): p. 471-493.